\documentclass[letterpaper,twocolumn,10pt]{article}
\usepackage{usenix-2020-09}

\usepackage{authblk}
\usepackage[english]{babel} 
\usepackage[utf8]{inputenc}

\usepackage[all]{nowidow}
\usepackage{fontawesome}
\usepackage{pdfcomment}

\usepackage{setspace}
\makeatletter
\renewcommand\AB@affilsepx{ \protect\Affilfont} 
\renewcommand\@fnsymbol[1]{
    \ifcase#1\or *\or †\or ‡\or §\or ‖\or ¶\else \@arabic{#1}\fi}
\makeatother
\usepackage{algpseudocode}
\usepackage[ruled,vlined,linesnumbered]{algorithm2e}
\usepackage{xcolor}
\usepackage{graphicx}
\usepackage{textcomp}
\usepackage{enumitem}
\usepackage{courier}
\usepackage{wrapfig,lipsum}

\usepackage[most]{tcolorbox}

\usepackage{xcolor,colortbl}
\usepackage[toc,page]{appendix}
\usepackage{booktabs}
\usepackage{multirow}
\usepackage{pifont}
\usepackage{bbding}
\usepackage{dingbat}
\usepackage{soul}
\usepackage{url}                
\usepackage{xcolor}             
\usepackage{url}
\usepackage{xcolor}
\usepackage{tcolorbox}
\newcommand{\tech}{\mbox{\textsc{{Appatch}}}}

\usepackage{breakurl}

\usepackage[available,functional]{usenixbadges}

\begin{document}
%
%
%

\title{{\tech}: 
Automated Adaptive Prompting Large Language Models for Real-World Software Vulnerability Patching
}

\author[1]{\rm Yu Nong}
\author[2]{\rm Haoran Yang}
\author[3]{\rm Long Cheng}
\author[1]{\rm Hongxin Hu}
\author[1\thanks{Haipeng Cai is the corresponding author.}]{\rm Haipeng Cai}

\affil[1]{\textit{University at Buffalo},}
\affil[2]{\textit{Washington State University},}
\affil[3]{\textit{Clemson University}}



\affil[ ]{\vspace{1.0em}\small
{\tt $^1$\{\textit{yunong,hongxinh,haipengc}\}\textit{@buffalo.edu}},
{\tt $^2$\textit{haoran.yang2@wsu.edu}},
{\tt $^3$\textit{lcheng2@clemson.edu}}
}

\maketitle

\begin{abstract}
Timely and effective vulnerability patching is essential for cybersecurity defense, for which various approaches have been proposed yet still struggle to generate valid and correct patches for real-world vulnerabilities. 
In this paper, we leverage the power and merits of pre-trained language language models (LLMs) to enable automated vulnerability patching using no test input/exploit evidence and without model training/fine-tuning.  
To elicit LLMs to effectively reason about vulnerable code behaviors, which is essential for quality patch generation, we introduce \textit{vulnerability semantics reasoning} and \textit{adaptive prompting} on LLMs and instantiate the methodology as {\tech}, an automated LLM-based patching system. 
Our evaluation of {\tech} on 97 zero-day vulnerabilities and 20 existing vulnerabilities demonstrates its superior performance to both existing prompting methods and state-of-the-art non-LLM-based techniques 
(by up to 28.33\% in F1 and 182.26\% in recall over the best baseline). 
Through {\tech}, we demonstrate what helps for LLM-based patching and how, as well as discussing what still lacks and why. 
\end{abstract}




\vspace{-0pt}
\section{Introduction}\label{sec:intro}
\vspace{-0pt}
Software vulnerabilities pose a constant and significant threat to the security of modern cyberspace~\cite{vulconsequence231,vulconsequence232} and they are widespread~\cite{cvedashboard23,fu2021flowdist,wen22usenixsecurity,wen23usenixsecurity,wen23ccs}. 
With the rapid growth in the volume and sophistication of cyberattacks, which are enabled by these vulnerabilities, timely and effective vulnerability patching has become increasingly critical~\cite{pearce2023examining,zhou2024largeb,wu2023effective}. 
However, manual approaches to patching are costly~\cite{vulcost23} and struggle to keep pace with the emergence of new vulnerabilities~\cite{iannone2022secret}. 
Thus, automated solutions are needed to help developers patch vulnerabilities before they are exploited~\cite{fu2022vulrepair,zhang2022program}.

In this context, various approaches to automated patching have been proposed~\cite{ma2017vurle,zhang2022example,gao2021beyond,chen2022neural,fu2022vulrepair,pearce2023examining}, which generally fall in two major categories: those based on code analysis and those based on deep learning (DL). 
Among other techniques, state-of-the-art (SOTA) code-analysis-based approaches achieve property-based repair based on access range analysis hence enforcing given safety properties~\cite{huang2019using} or 
use sanitizers to extract vulnerability-representing constraints for symbolic-execution-based repair~\cite{gao2021beyond}. More recently, inductive property inference along with simple templates~\cite{zhang2022program} is employed to generate vulnerability patches. 
However, these techniques rely on exploits or vulnerability-triggering test inputs, which may not always be available, especially for emerging (e.g., zero-day) vulnerabilities. They often also require that the code to patch be compilable, which further limits their applicability (e.g., they can not be used before the project becomes complete). 

DL-based approaches overcome these limitations and have gained more momentum in recent years~\cite{wu2023effective,fu2022vulrepair,chen2022neural}. These techniques leverage existing vulnerability datasets to train a DL model that learns known vulnerability-fixing patterns, hence the ability to repair a given vulnerability. 
Despite their promises, they need sizable and quality training data with vulnerability labels, which is not widely and diversely available~\cite{nong2022generating,bhandari2021cvefixes,nong2022open}. 
As a result, they do not generalize well to unseen code and often stumble on real-world vulnerabilities~\cite{wu2023effective}. 
While data augmentation helps~\cite{nongvulgen,nongvgx,cai2023generating}, recent studies show that the augmented models still suffer low accuracy~\cite{nongvgx,nong2024chain}. 


Most recently, large language models (LLMs) have emerged as powerful tools to assist developers with coding related tasks. LLMs overcome the lack of generalizability of ordinary DL-based approaches and have shown promising potential for vulnerability patching~\cite{pearce2023examining,nong2024chain}. 
However, the SOTA LLM-based (with zero-shot) technique~\cite{pearce2023examining} does not work well on real-world vulnerabilities and often generates no patches or invalid ones~\cite{nong2024chain}. 
The recent study~\cite{nong2024chain} shows that LLM can be significantly improved by carefully designed prompting strategies, especially based on chain-of-thought (CoT)~\cite{wei2022chain} reasoning. 
Prompting is indeed attractive in that it does not require (pre-)training or even fine-tuning, which would not only require sizable labeled datasets but also be infeasible to ordinary users, especially on foundation-scale LLMs that may actually hold the promise for effective vulnerability analysis. 

Despite a promising direction, vulnerability patching via prompting LLMs faces several critical challenges as revealed lately~\cite{nong2024chain,wu2023effective,nong2024automated}. 
First, for automated patching, the prompting would need to be automated too. Yet it is not known how to automatically write a prompt to an LLM for vulnerability patching (\textit{Challenge 1}). 
Second, prompting tends to include exemplars to be effective, especially if it aims to elicit reasoning on LLMs~\cite{wei2022chain} as CoT does, which is essential for vulnerability patching---a task relying on code semantics reasoning. 
Yet again, for an arbitrary program to patch, how to prepare the best exemplars is unknown (\textit{Challenge 2}). 
Third, LLMs are known to suffer from capacity (\#tokens) and context constraints, which 
makes it difficult to provide necessary code context in a prompt and analyze sizable code; yet vulnerabilities are context-sensitive and can be interprocedural, thus patching them often requires accommodating potentially larger code contexts (\textit{Challenge 3}).
Moreover, LLMs are commonly subject to hallucination and non-deterministic responses, whereas stably generating valid and correct patches is essential for  
vulnerability patching---bad patching can be worse than not patching (\textit{Challenge 4}).

In this paper, we present {\tech}, a novel, automated \ul{A}daptive \ul{p}rompting methodology for LLM-based vulnerability \ul{patch}ing. 
Given a vulnerable program with a known vulnerability location, 
{\tech} first narrows down the scope of analysis to only the relevant subset of the program via a step called \textit{semantics-aware scoping}. This allows LLMs to analyze a smaller code snippet that includes all the essential information necessary for reasoning about the vulnerability-related code behaviors, addressing \textit{Challenge 3}. 
Then, it elicits the LLM to identify the vulnerability's root cause within the reduced scope with vulnerability semantics reasoning, henceforth selecting exemplars that best fit the program on the fly from a pre-mined exemplar database, an essential step referred to as dynamic \textit{adaptive prompting}, which addresses \textit{Challenges 2}. 
The database is built offline by mining exemplars from known patches. 
With the adaptively chosen exemplars, {\tech} forms the patching prompt automatically to generate multiple candidate patches with LLMs, addressing \textit{Challenge 1}. 
To mitigate model hallucinations, {\tech} consults an ensemble of LLMs to cross-validate the candidate patches concerning multiple facets of patch quality---validity (not changing functionality) and correct (fixing the vulnerability), hence addressing \textit{Challenge 4}. 

We have implemented {\tech} as an open-source tool based on four latest and most powerful LLMs: GPT-4, Gemini-1.5, Claude-3.5, and Llama-3.1, and evaluated it against a dataset with 97 zero-day (across 18 unique projects) samples and a dataset with 20 existing real-world vulnerability samples. 
With {\tech}, the evaluated LLMs achieve up to 36.46\% and 73.86\% F1 on the two datasets while the same LLMs with baseline prompting strategies only achieve up to 28.41\% and 68.54\% F1, respectively. {\tech} also outperforms SOTA traditional vulnerability patching approaches, where they only achieve up to 17.53\% and 85.00\% recall on the two datasets while {\tech} achieves 49.48\% and 90.00\% recall, respectively. {\tech} is also reasonably efficient, taking on average 37.148-50.209 seconds to generate a patch, and using 5,684-6,802 tokens per patch with the LLMs. 

Our results suggest the practical feasibility and merits of LLM-based vulnerability patching, via a novel methodology (pivoted by \textit{automated adaptive prompting based on vulnerability semantics reasoning}) of leveraging LLMs to generate valid and correct patches. 
Through {\tech}, we demonstrated how LLMs can be helped for this challenging task; also, by examining failure cases, we discuss how/where and why LLMs may still fall short even with {\tech}.

\begin{figure}[tp]
\centering
\vspace{5pt}
	\includegraphics[width=0.90\linewidth]{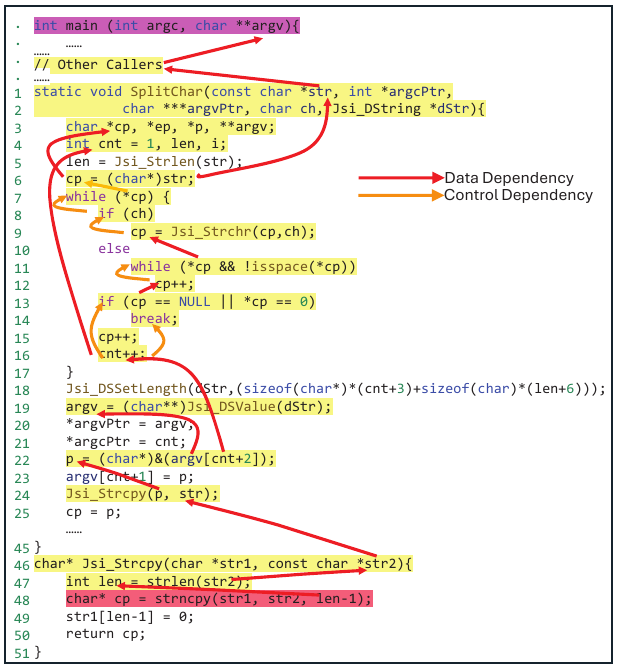}
        \vspace{-10pt}
        \caption{An example vulnerable program sample where the vulnerable statement is at line 48.}
	\label{fig:example}
        \vspace{-2pt}
\end{figure}

\begin{figure}[tp]
\centering
\vspace{2pt}
	\includegraphics[width=0.90\linewidth]{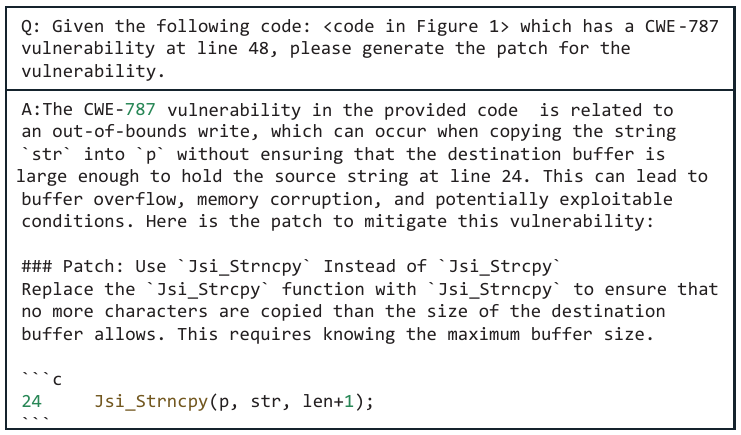}
        \vspace{-10pt}
        \caption{GPT-4's patch for the program sample in Figure~\ref{fig:example} with \textit{standard prompting}.}
	\label{fig:discussion-std}
        \vspace{-2pt}
\end{figure}

\begin{figure}[tp]
\centering
\vspace{4pt}
	\includegraphics[width=0.80\linewidth]{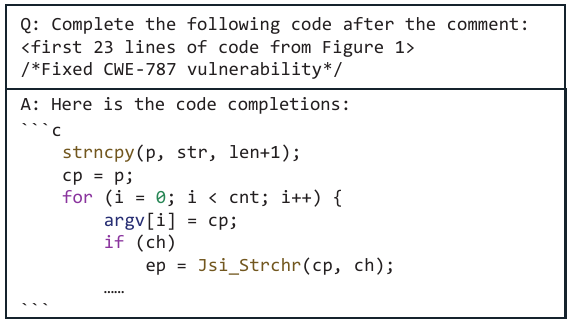}
        \vspace{-8pt}
        \caption{GPT-4's patch for the sample in Figure~\ref{fig:example} with a state-of-the-art LLM-based approach \textit{zero-shot completion}.}
	\label{fig:discussion-s2}
        \vspace{-2pt}
\end{figure}

\vspace{-0pt}
\section{Background and Motivation}
\vspace{-0pt}
LLMs have been explored for vulnerability analysis tasks such as vulnerability detection~\cite{noever2023can}, patching~\cite{pearce2023examining, ullah2024llms}, and secure code generation~\cite{he2023controlling}. Among existing approaches to vulnerability patching, a popular method is to use a standard prompting strategy, which directly asks the LLMs to patch the vulnerable code~\cite{ullah2024llms}. However, this method does not provide necessary guidance to LLMs, leaving all the vulnerability analysis reasoning process to the models. Another, SOTA method is zero-shot code completion~\cite{pearce2023examining}, which removes the vulnerable code and lets LLMs complete the vulnerable parts. However, this approach treats vulnerability patching as a simple code completion task, without code analysis/reasoning. 

To show the limitations of existing LLM-based vulnerability patching approaches, we test the two methods mentioned above on the example in Figure~\ref{fig:example}. As shown, the example has a CWE-787 
vulnerability at line 48. The vulnerability is caused by the possible out-of-bound write to the pointer {\tt str1} as the string {\tt str2} may be longer than the boundary of (the memory region pointed to by) {\tt str1}. Figure~\ref{fig:discussion-std} shows GPT-4's output patch with standard prompting, where we directly ask the model to patch the vulnerability with the vulnerable line and CWE ID provided. 
As seen, GPT-4 seems to recognize 
that the vulnerability is caused by the insufficient size of the 
memory $p$ points to at line 24
and it should ensure that the write is not longer than {\tt p}'s boundary. While the basic idea is correct, it is also needed to find out how much memory {\tt p} should point to for patching the vulnerability. In this case, necessary data/control dependency analysis is needed. However, GPT-4 generates the patch by changing {\tt Jsi\_Strcpy} to {\tt Jsi\_Strncpy} and adding a write limit {\tt len+1} without analyzing the code semantics. Thus, it fails to patch the vulnerability.

Figure~\ref{fig:discussion-s2} shows GPT-4's output patch with the zero-shot code completion method. As shown, we provide the code up to line 23 (where the ground-truth patch is located) and prompt the model (via a comment) to complete the remaining code while addressing the CWE-787 vulnerability. Similar to standard prompting, it simply completes line 24 with {\tt strncpy} and sets a write limit {\tt len+1} without analyzing the actual size of {\tt p}. Thus, it also fails to patch the vulnerability. 

Based on the two examples above, it is clear that, to patch real-world vulnerabilities effectively, even powerful LLMs may still need to be guided step by step for necessary reasoning about code semantics.  
%
Therefore, we introduce {\tech}, an automated framework that adaptively prompts LLMs for vulnerability patching with semantics reasoning guidance.
We introduce four key design elements in {\tech}:
\begin{enumerate}[itemsep=0pt,topsep=2pt]
    \item To help LLMs concentrate on the most essential parts of a program for vulnerability patching, we introduce semantics-aware scoping, which only provides LLMs with code entities that capture the core vulnerable behavior of the program (noted as \textit{vulnerability semantics}).
    \item To guide LLMs to analyze and patch vulnerabilities correctly, we arm them with vulnerability semantics reasoning capabilities via CoT prompting~\cite{wei2022chain} that  demonstrates correct reasoning steps with a few exemplars.
    \item Considering the diversity of vulnerability root causes and patching strategies, a large-scale exemplar pool is needed. However, manually writing the reasoning steps for the exemplars is time-consuming. Thus, we design an automated exemplar mining module which generates exemplars based on existing real-world samples. 
    \item To provide the best-fit exemplars for a given testing sample (i.e., code to patch), we design dynamic adaptive prompting which automatically picks exemplars based on the vulnerability root causes of the testing sample.
\end{enumerate}




\vspace{-0pt}
\section{Technical Design}
\vspace{-0pt}
Now we present our technical approach, starting with a design overview followed by details on each component. 

\vspace{-0pt}
\subsection{Overview}
\vspace{-0pt}
\begin{figure*}[tp]
\centering
\vspace{2pt}
	\includegraphics[width=0.99\linewidth]{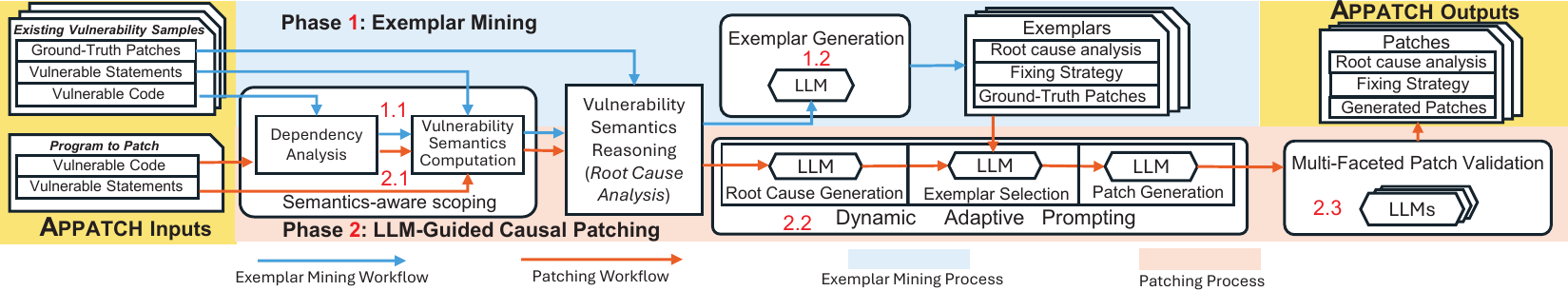}
        \vspace{-4pt}
        \caption{An overview of {\tech}'s design, including its inputs, two main working phases, and outputs.}
	\label{fig:overfiew}
        \vspace{4pt}
\end{figure*}

The overall design of {\tech} is depicted in Figure~\ref{fig:overfiew}. 
We target a realistic vulnerability patching scenario with {\tech}, in which 
(1) the vulnerability manifestation locations (i.e., \textit{vulnerable statements}) and (2) the vulnerability types (i.e., \textit{CWE IDs}) of a given program to patch are available. 
Prior to the patching time, {\tech} demonstrates to LLMs correct reasoning towards patch generation with (3) existing vulnerable samples for which the ground-truth patches are known in addition. 
These three constitute the \textbf{inputs} to {\tech}.

With these inputs, {\tech} operates in two phases. 
In \textbf{Phase 1: Exemplar Mining}, through \textit{semantics-aware scoping} (\textbf{1.1}), it union-slices~\cite{beszedes2002union} each given vulnerable sample to capture its \textit{vulnerability semantics} (essential behavior of the vulnerability, as formally defined in $\S$\ref{sec:s3.2}). Then, 
{\tech} proceeds with \textit{exemplar generation} (\textbf{1.2}), producing a pool of exemplars each including the root cause analysis, fixing strategy, and ground-truth patch. The root cause and fixing strategy are also generated by the LLMs automatically, according to the vulnerability semantics \textit{along with the ground-truth patch}. 

Next, in \textbf{Phase 2: LLM-Guided Causal Patching}, the given program to patch is also scoped into vulnerability semantics (\textbf{2.1}). Then, through \textit{dynamic adaptive prompting} (\textbf{2.2}), {\tech} analyzes the root cause of the given vulnerability with an LLM and selects similar exemplars based on the root cause. With the chosen exemplars, candidate patches are generated via CoT prompting on the LLM. 
Finally, the generated patches are validated and refined through multiple LLMs through a \textit{multi-faceted patch validation} (\textbf{2.3}) process. 

As a result, the validated patches are returned as the \textbf{outputs} of {\tech}. 
Note that while both phases rely on LLMs for automated root cause and fixing strategy identification using the vulnerability semantics reasoning, Phase 1 additionally feeds the ground-truth patches to the models, This addition is essential because it is expected to help LLMs generate root cause and fixing strategy that are more likely correct than otherwise hence serving the demonstrating (exemplar) purposes to elicit LLMs' similar reasoning at patching time.  

\vspace{-0pt}
\subsection{Vulnerability Semantics}\label{sec:s3.2}
\vspace{-0pt}
We introduce this core concept underlying {\tech} with following insights. 
First, although a vulnerable program may contain many lines of code, oftentimes only a relatively small portion of those lines actually causes the vulnerability. 
By focusing on this portion, 
which scopes the semantics of the code that makes it vulnerable (intuitively noted as \textit{vulnerability semantics} for now)
we can enable LLMs to perform more effective vulnerability analysis.
Second, mainstream LLMs often struggle with processing and generating long texts. 
By narrowing the scope to vulnerability semantics, we can also reduce the length of the exemplars, prompts, and responses,
thereby minimizing potential distractions for the LLMs and improving their performance. Third, 
vulnerability-relevant control/data flow is what is difficult for LLMs to understand without guidance on reasoning about the flow~\cite{purba2023software}---thus, vulnerability semantics-based reasoning is necessary to guide LLMs to conduct effective vulnerability patching.

Therefore, we define vulnerability semantics as the aspects of a program's code that contribute to its vulnerabilities. These of course include statements where the vulnerabilities are triggered. However, such statements alone do not cause vulnerabilities unless they can be exploited via external inputs. Thus, to effectively capture vulnerability semantics, it is important to consider how these statements are influenced by the program's external inputs through control and data flow dependencies. Therefore, we also include the contextual statements that influence the vulnerability statements as part of vulnerability semantics, based on the data and control flow. 


Given a program \( P \), which consists of a set \( S \) of statements, the set of \textbf{vulnerability locations} \( S_v \subseteq S \) are statements where the vulnerability manifests. The set of \textbf{external inputs} \( EI \subseteq S \) consists of statements where external input data is received or retrieved by \( P \). This includes program input variables, returns from external function calls that read data, and global states modified by other functions (such as memory allocations, file input, network input, etc.). The \textbf{vulnerability semantics} \( \mathcal{V}(P, S_v, EI) \) is defined as the union of all \textbf{static backward slices} of the program \( P \) starting from any of the vulnerability locations \( S_v \) and terminating at any of the external inputs in \( EI \). That is, 
\[ \mathcal{V}(P, S_v, EI) = \bigcup_{e_i \in EI}~\bigcup_{s_v \in S_v}\text{Slice}(P, s_v, e_i) \] where:

\begin{itemize} [leftmargin=*,itemsep=0pt, topsep=0pt]
\item \( \text{Slice}(P, s_v, e_i) \) is the set of all statements in the static backward slice from \( s_v \) to \( e_i \). 
\item 

A \textbf{static backward slice} from \( s_v \) to \( e_i \) is the subset of statements in \( P \) that  
affect the execution at \( s_v \), tracing backward along control and data dependencies to reach \( e_i \).

\end{itemize} 
Note that the vulnerability locations and external inputs may be in different functions. Therefore, we perform \emph{interprocedural} backward slicing to capture the  vulnerability semantics comprehensively. In this way, \( \mathcal{V}(P, S_v, EI) \) represents the set of all statements in \( P \) that are relevant to the flow of data and control from any of the external inputs \( EI \) to any of the vulnerability locations \( S_v \).

As an example, Figure~\ref{fig:example} illustrates a partial representation of vulnerability semantics. We start from a vulnerable statement (\( s_v \)) at line 48 and conduct union slicing to the program input based on data and control dependencies (marked with \textcolor{red}{red} and \textcolor{orange}{orange} arrows). As a result, the statements marked as {\textcolor[rgb]{0.8,0.8,0}{yellow}} are the contexts of the vulnerable statement. The vulnerable statement (marked in {\textcolor{red}{red}}), the external input (in {\textcolor{purple}{purple}}), and its contextual statements constitute (the \textit{code representation} of) the vulnerability semantics for the sample.


\vspace{-0pt}
\subsection{Exemplar Mining (Phase 1)}
\vspace{-0pt}
\subsubsection{Semantics-Aware Scoping (Step 1.1)} \label{sec:s1.1}
\vspace{-0pt}
\setlength{\dbltextfloatsep}{6pt}
\begin{algorithm}[tp]
\scriptsize
\caption{Semantics-Aware Scoping}
\label{algo:semanticsscoped}
\SetKwProg{Fn}{Function}{}{end}
\SetKwFunction{SemanticsAwareScoping}{SemanticsAwareScoping}
\SetKwFunction{ExtractSDG}{ExtractSDG}
\SetKwFunction{IdentifyExternalInputs}{IdentifyExternalInputs}
\SetKwFunction{ComputeBackwardSlice}{ComputeBackwardSlice}
\SetKwFunction{InitializeSlice}{InitializeSlice}
\SetKw{Return}{return}
\LinesNumbered
\KwIn{\( P \): input program, \( S_v \): vulnerable statements
}
\KwOut{\( \mathcal{V}(P, S_v, EI) \): vulnerability semantics (code representation)}

\Fn{\SemanticsAwareScoping{\(P, S_v, EI\)}}{
    $sdg$ $\gets$ \ExtractSDG{\(P\)};\hfill {\color{magenta}\tiny // Construct SDG for the program $P$}\\
    \(EI\) $\gets$ \IdentifyExternalInputs{\(P\)}; \hfill {\color{magenta}\tiny // Identify external inputs} \\
    \(\mathcal{V}\) $\gets$ \InitializeSlice{};\hfill {\color{magenta}\tiny // Initialize an empty set} \\
    \ForEach{\(e_i\) in \(EI\)} {
        \ForEach{\(s_v\) in \(S_v\)}{
            \(BS_{ei} \gets \) \ComputeBackwardSlice{\(sdg, s_v, e_i\)}; \hfill {\color{magenta}\tiny // Backward slice starting from a vulnerable location $s_v$ until reaching an external input $e_i$} \\ 
            \(\mathcal{V} \gets \mathcal{V} \cup BS_{ei} \); \hfill {\tiny \color{magenta}// Unionize the above backward slice}
        }
    }
  
    \Return \(\mathcal{V}\)\;
}
\end{algorithm}
\vspace{-0pt}



In this step, {\tech} extracts the essential code 
that represents the vulnerability semantics
from each of the existing vulnerable code samples.
To that end, 
it first analyzes the sample and constructs its system dependence graph (SDG)~\cite{horwitz1990interprocedural} through an \textit{interprocedural dependence analysis}. 
An SDG is a representation that combines both data dependencies and control dependencies of a program; thus, 
it is essential in {computing the vulnerability semantics} of a program.
As shown in Algorithm~\ref{algo:semanticsscoped}, the \textit{vulnerability semantics computation} begins this process by building the SDG (line 2). Then, we perform a simple static analysis to realize \texttt{IdentifyExternalInputs} (line 3). We first curate a set of external functions that may produce external inputs or cause side effects to program states (e.g., malloc, socket\_recv, scanf). Then, it traverses the program's control flow graph (CFG), identifying each callsite of any of these external functions. To be more comprehensive, it also considers the definition sites of program’s input variables as an external input (we simply treat the program entry point as the joint entity of those definition sites).

As the main loop begins, we compute the interprocedural backward slice from each of the vulnerable statements \(s_v\) to each of the external input \(e_i\) (lines 5-8), via another static analysis based on the SDG just constructed. 
It traverses the $sdg$ from $s_v$, identifies backward dependencies transitively until $e_i$ is reached (which has no further backward dependence). 
Finally, we 
compute the union slice~\cite{beszedes2002union} for all external inputs in $EI$ (line 8) and return it 
as the (code representation of the) vulnerability semantics (line 9).

\vspace{-0pt}
\subsubsection{Exemplar Generation (Step 1.2)}
\vspace{-0pt}
With the vulnerability semantics computed in Step 1.1, Step 1.2 generates the exemplars 
from the given existing vulnerable samples. 
Since our dynamic adaptive prompting leverages CoT prompting which guides LLMs to generate quality patches with appropriate intermediate reasoning, we 
divide the overall reasoning process into three high-level steps: (1) \textit{vulnerability semantics reasoning (as root cause analysis)}, (2) fixing strategy identification, and (3) patch generation. However, it is difficult to write the reasoning for each collected sample manually. Therefore, we leverage LLMs to automatically generate the exemplars. Since we have the ground-truth patches in the given existing samples, we prompt the LLMs to generate (1) 
with the following template:
\vspace{-6pt}
\begin{tcolorbox}[left=1pt,right=1pt,top=0pt, bottom=3pt, lifted shadow={1mm}{1mm}{1mm}{1mm}{black!100!white}, colback=white, arc=0pt,auto outer arc, boxrule=.5pt]
\textbf{Q}: Given the following code slice $\mathcal{V}_{exem}$, which has a vulnerability among $<$CWE-IDs$>$ and lines $S_v$, the patch is $<$ground-truth patch$>$.
\emph{Starting with the external inputs: <$EI$ identified>, reason about the vulnerable behavior step by step until the vulnerability is determined. }
\vspace{-4pt}
\end{tcolorbox}
\vspace{-4pt}

Since the code slice $\mathcal{V}$ is interprocedural across the whole program, it is likely to contain a large number of code statements~\cite{binkley2007empirical}, making it difficult for LLMs to process. Therefore, the \textit{EI identified} are only part of the full set of EI identified in Algorithm~\ref{algo:semanticsscoped} that reach any of the patch locations. Then, the code slice provided to LLMs $\mathcal{V}_{exem}$ only contains the paths between any of such identified EI subset and any of 
the vulnerable statements.
This helps LLMs concentrate on the most essential parts of the program for vulnerability patching.

\begin{figure*}[tp]
\centering
	\includegraphics[width=0.8\linewidth]{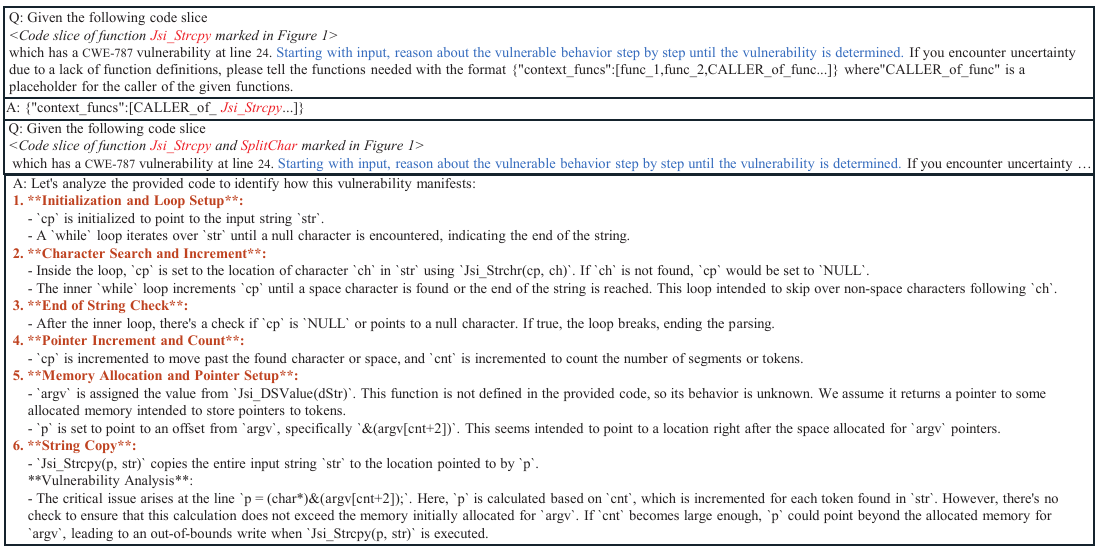}
        \vspace{-12pt}
        \caption{Root cause analysis generated by GPT-4 \emph{\textbf{with} vulnerability semantics reasoning} for the sample in Figure~\ref{fig:example}.}
	\label{fig:example-cause-sem}
        \vspace{-0pt}
\end{figure*}

In this template (as well as the one for dynamic adaptive prompting's root cause generation), we leverage vulnerability-semantics-guided reasoning, which is described as "\emph{Starting with the external inputs, reason about the vulnerable behavior step by step until the vulnerability is determined}". Although this is simple, it significantly helps LLMs analyze the vulnerability root cause comprehensively and effectively. Figures~\ref{fig:example-cause-sem} and \ref{fig:example-cause-nosem} show the comparison between the root cause analysis with and without vulnerability-semantics-guided reasoning. Without such reasoning, the output misses many details for the vulnerability causes, which make the analysis inaccurate. 

The key insight underlying our 
automated exemplar generation is that, with the ground-truth patches provided, LLMs are capable of generating the correct reasoning steps which are helpful to guide themselves to generate patches given new samples to patch. To corroborate this hypothesis, we conduct a preliminary experiment on the real-world samples from our collected dataset (see $\S$\ref{ssec:rq1}). Based on our manual inspection on the generated exemplars by GPT-4, 92.98\% of the reasoning are correct, indicating that using LLMs to automatically generate exemplars is effective. These exemplars compose the quality exemplar pool for the dynamic adaptive prompting step (Step 2.2) to select exemplars from. 


\vspace{-0pt}
\subsection{LLM-Guided Causal Patching (Phase 2)}
\vspace{-0pt}
\subsubsection{Semantics-Aware Scoping (Step 2.1)}
\vspace{-0pt}
In Phase 2, given a testing sample that contains vulnerable code and its vulnerability location, {\tech} again computes the vulnerability semantics with the semantics-aware scoping module. The process is the same as the one in Step 1.1: we construct the SDG from the source code, then get the vulnerability semantics via 
Algorithm~\ref{algo:semanticsscoped}. 
\vspace{-0pt}
\subsubsection{Dynamic Adaptive Prompting (Step 2.2)}
\vspace{-0pt}
With the vulnerability semantics computed in Step 2.1, we perform dynamic adaptive prompting to guide the LLMs to generate patches. 
The overall algorithm of this dynamic adaptive prompting step 
is shown in Algorithm~\ref{algo:dynamic}. {\tech} leverages a \emph{progressive prompting} strategy which prompts the LLMs to generate the root cause, select exemplars, and generate the patch in separated prompts. 
This is to (1) guide the LLMs to generate the patches step by step and (2) allow LLMs to obtain the information and exemplars on demand. 

\begin{algorithm}[h]
\scriptsize
\caption{Dynamic Adaptive Prompting}
\label{algo:dynamic}
\SetKwProg{Fn}{Function}{}{end}
\SetKwFunction{DynPrompting}{DynPrompting}
\SetKwFunction{GenCause}{GenCause}
\SetKwFunction{GetVulFuncSlice}{GetVulFuncSlice}
\SetKwFunction{SizeOf}{SizeOf}
\SetKwFunction{GetExemCause}{GetExemCause}
\SetKwFunction{CompareCause}{CompareCause}
\SetKwFunction{GenPatch}{GenPatch}
\SetKwFunction{GenCauseByLLM}{GenCauseByLLM}
\SetKwFunction{GenCauseByLLM}{GenCauseByLLM}
\SetKwFunction{HasExpandDemand}{HasExpandDemand}
\SetKwFunction{GetDemandedFuncSlice}{GetDemandedFuncSlice}
\SetKw{Return}{return}
\SetKw{Break}{break}
\LinesNumbered
\KwIn{\emph{vul\_slice}: vulnerable slice, \emph{exemplars}: mined exemplars}
\KwOut{\emph{cand\_patches}: candidate patches}

\Fn{\DynPrompting{\emph{vul\_slice}, \emph{exemplars}}}{
    \emph{cause} $\gets$ \GenCause{\emph{vul\_slice}}\hfill {\color{magenta}\tiny // Generate root cause of the slice} \\
    \textit{chosen\_exemplars} $\gets$ []\hfill {\color{magenta}\tiny // Initialize an empty list}\\
    \ForEach{$exemplar$ in Exemplars}{
        \textit{exemplar\_cause} $\gets$ \GetExemCause{exemplar}\hfill{\color{magenta}\tiny // Extract cause}\\
        \textit{comp\_ans} $\gets$ \CompareCause{cause, exemplar\_cause}\\
        \If{\textit{comp\_ans} is Yes}{
            \textit{chosen\_exemplars}.insert(\textit{exemplar})\hfill{\color{magenta}\tiny // Add similar exemplars}\\ 
        }
        \If{\SizeOf{chosen\_exemplars}$\ge$8}{
            \Break \hfill {\color{magenta}\tiny // Break when the exemplars are enough}
        }
    }
    \textit{cand\_patches}$\gets$\GenPatch{chosen\_exemplars, vul\_slice} \\
    \Return \textit{cand\_patches}\;
}
\Fn{\GenCause{\emph{vul\_slice}}}{
    \textit{cause\_slice}$\gets$\GetVulFuncSlice{vul\_slice} \hfill {\color{magenta}\tiny// Get vulnerable function slice} \\
    \textit{LLMAns} $\gets$ \GenCauseByLLM{cause\_slice} \hfill {\color{magenta}\tiny// Get response from LLMs} \\
    \While{
         \HasExpandDemand{\textit{LLMAns}} 
    }{
        \textit{cause\_slice} $\gets$ \textit{cause\_slice} $\cup$ \GetDemandedFuncSlice{\textit{LLMAns}} \\
        \textit{LLMAns} $\gets$ \GenCauseByLLM{cause\_slice} \\
    }
    \Return \textit{LLMAns} \\
}
\end{algorithm}

\textbf{Root Cause Generation}. 
We again leverage vulnerability semantics reasoning to generate root cause analysis:
given the vulnerability semantics computed (i.e., the union slice noted as $\mathcal{V}_{cause}$), the CWE IDs, and the vulnerable statements $S_v$, {\tech} first prompts the LLMs to generate the root cause (i.e., the natural-language description of the vulnerability semantics reasoning steps). 
We prompt the LLM with the following template:
\vspace{-6pt}
\begin{tcolorbox}[left=1pt,right=1pt,top=0pt, bottom=3pt, lifted shadow={1mm}{1mm}{1mm}{1mm}{black!100!white}, colback=white, arc=0pt,auto outer arc, boxrule=.5pt]
 \textbf{Q}: Given the following code slice: $\mathcal{V}_{cause}$ which has a vulnerability among $<$CWE-IDs$>$ and lines: $S_v$. \emph{Starting with the external inputs: <$EI$ identified>, reason about the vulnerable behavior step by step until the vulnerability is determined. } If you encounter uncertainty due to a lack of function definitions, please tell the functions needed with the format \{"context\_funcs":[func\_1,func\_2,CALLER\_of\_func...]\} where "CALLER\_of\_func" is a placeholder for the caller of the given functions.

\vspace{-4pt}
\end{tcolorbox}
\vspace{-4pt}

\begin{figure}[h]
\centering
	\includegraphics[width=0.86\linewidth]{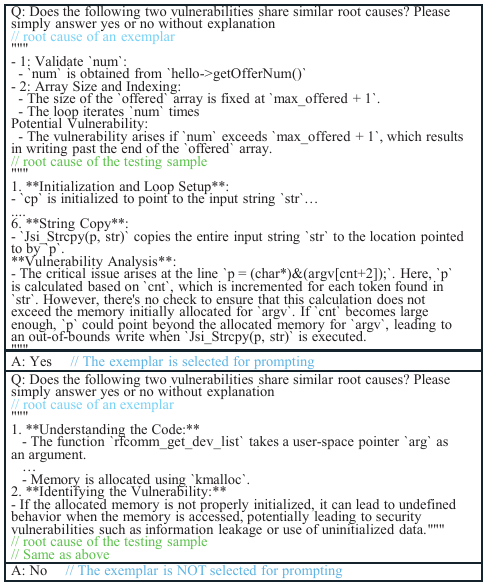}
        \vspace{-12pt}
        \caption{Exemplar selection by GPT-4 for the sample in Figure~\ref{fig:example}.}
	\label{fig:example-select}
        \vspace{-6pt}
\end{figure}

In the prompt above, we start from the slice of the function involving the vulnerable statements (line 14 in Algorithm~\ref{algo:dynamic}) and expand the slice on LLMs' demand. Note that the EI identified are only the part of EI identified in Algorithm I that are relevant to the functions the LLM asked for so far.  
The rationale is that LLMs usually have input token limitation and thus we cannot input the whole projects into the models. Even if the models are capable of accepting more tokens, the irrelevant functions would distract the LLMs in analyzing the vulnerabilities. Therefore, we only provide the part of the slice that involves the functions that the LLM asks for. This progressive prompting scheme, similar to LLift~\cite{li2024enhancing}, 
is essential for {\tech} to achieve \textit{interprocedural} vulnerability analysis and patching. 

The algorithm for generating the root cause is shown in Algorithm~\ref{algo:dynamic} lines 13-19. Given the vulnerability code slice which may contain a large number of functions and statements, we start with the part of the slice only involving the functions that contain the vulnerability manifestation statements (line 14). Then, we iteratively query LLMs with the prompt template given earlier until the LLMs do not ask for additional functions (lines 15-18). The final root cause analysis is then returned to the main routine (\textit{DynPrompting}) for dynamic adaptive prompting.




\textbf{Exemplar Selection.} 
After generating the root cause analysis for the testing sample, we select appropriate exemplars to prompt the LLMs to generate the patches. 
As shown in Algorithm~\ref{algo:dynamic}, an empty list, \textit{chosen\_exemplars}, is initialized to store exemplars that closely match the root cause (line 3). The selection process iterates through the set of pre-mined exemplars and gets the vulnerability root causes from them (lines 4-5). 
To select the exemplars, we prompt the LLMs to compare the root cause analysis result of the testing sample and each exemplar sample (line 6) with the following prompt:
\vspace{-12pt}
\begin{tcolorbox}[left=1pt,right=1pt,top=0pt, bottom=3pt, lifted shadow={1mm}{1mm}{1mm}{1mm}{black!100!white}, colback=white, arc=0pt,auto outer arc, boxrule=.5pt]
    \textbf{Q}: 
    Are the following two root causes similar? 
    
    $<$exemplar\_sample\_root\_cause$>$
    
    $<$testing\_sample\_root\_cause$>$
    
    Please simply answer yes or no.

\vspace{-4pt}
\end{tcolorbox}
\vspace{-4pt}

\noindent
Considering that the complexity is $O(mn)$ where $m$ is the number of exemplar samples and $n$ is that of testing samples, we let LLMs directly answer yes or no to save resources. 

\begin{figure}[h]
\centering
	\includegraphics[width=0.9\linewidth]{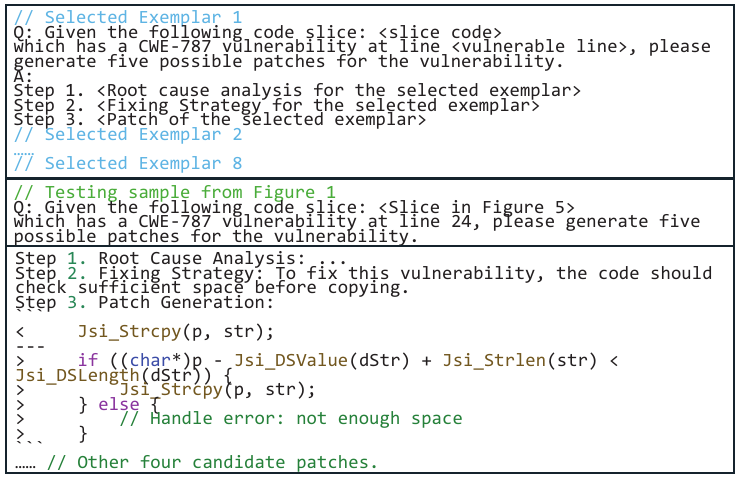}
        \vspace{-12pt}
        \caption{GPT-4's patch for the sample in Figure~\ref{fig:example} with {\tech}.}
	\label{fig:discussion-appatch}
        \vspace{-6pt}
\end{figure}

Figure~\ref{fig:example-select} shows parts of the exemplar selection process for the example in Figure~\ref{fig:example}. Once the used LLM answers yes, the root cause analysis, fixing strategy, and the ground-truth patch of that exemplar sample will be used as the exemplar to prompt the LLMs to patch the testing sample (lines 7-8). We follow the original CoT work~\cite{wei2022chain} and only select up to 8 exemplars for each testing sample. 

\textbf{Patch Generation.}
Once the exemplars are selected, {\tech} prompts the LLMs to generate the patch (line 11) with the following template:
\vspace{-6pt}
\begin{tcolorbox}[left=1pt,right=1pt,top=0pt, bottom=3pt, lifted shadow={1mm}{1mm}{1mm}{1mm}{black!100!white}, colback=white, arc=0pt,auto outer arc, boxrule=.5pt]
 $<$selected exemplars$>$
    
    \textbf{Q}: Given the following code slice: $\mathcal{V}_{cause}$ which has a vulnerability among $<$CWE-IDs$>$ and lines: $S_v$, please generate five possible patches for the vulnerability. 
    
    \textbf{A}: Step 1. $<$root cause analysis$>$

\vspace{-4pt}
\end{tcolorbox}
\vspace{-4pt}

In this prompt, we first provide the selected exemplars where each exemplar includes its vulnerability semantics (union slice), CWE-IDs, vulnerable statements, as well as the root cause analysis, the fixing strategy, and the ground-truth patch. This is to guide the LLMs to patch the testing sample with a similar, complete workflow. Then, we provide the testing sample's vulnerability semantics slice ($\mathcal{V}_{cause}$) used for the final root cause generation, CWE-IDs, and the vulnerable statements ($S_v$) with the same format as the exemplars. We also prompt the LLMs to generate more than one candidate patch so that the recall can be improved and developers have more choices to select appropriate patches from. To avoid repeated output, we also \textit{add the root cause generated in the previous step} 
for the LLMs to follow through. 
With the dynamically selected exemplars, the LLMs are expected to generate quality patches.

\begin{algorithm}[htp]
\scriptsize
\caption{Multi-Faceted Patch Validation}
\label{algo:validation}
\SetKwProg{Fn}{Function}{}{end}
\SetKwFunction{PatchValidation}{PatchValidation}
\SetKwFunction{ValPatchLLM}{ValPatchLLM}
\SetKw{Return}{return}
\SetKw{Break}{break}
\LinesNumbered
\KwIn{\emph{cand\_patches}: candatate patches, \emph{llms}: LLMs used for validation}
\KwOut{\emph{valid\_patches}: validated patches}
\Fn{\PatchValidation{\emph{cand\_patches}, \emph{llms}}}{
    \emph{valid\_patches} $gets$ \emph{cand\_patches} \\
    \ForEach{\emph{cand\_patch} in \emph{cand\_patches}}{ 
        retain $\gets$ False \\
        \ForEach{llm in llms}{ 
            \emph{val\_ans} $\gets$ \ValPatchLLM{cand\_patch, llm} {\color{magenta}\tiny // Validate the patch with an LLM}\\
            \If{\emph{val\_ans} is Yes}{
                retain$\gets$True {\color{magenta}\tiny // Remove the patch if all the LLMs think it is invalid} 
            }
        }
        \If{retain is False}{
            \emph{valid\_patches}.remove(\emph{cand\_patch}) 
        }
    }
    \Return \emph{valid\_patches}
}
\end{algorithm}

\vspace{-0pt}
\subsubsection{Multi-Faceted Patch Validation (Step 2.3)}
\vspace{-0pt}
After generating the candidate patches, we further leverage multi-faceted (i.e., fixing vulnerability and preserving functionality) patch validation to reduce incorrect patches which are not useful for developers. Considering that the same LLM may not be effective for validating the generated patches by itself, we leverage the ensemble method to conduct the validation. Therefore, multiple effective LLMs for vulnerability patching are used for the validation. 

\begin{figure}[tp]
\centering
	\includegraphics[width=0.99\linewidth]{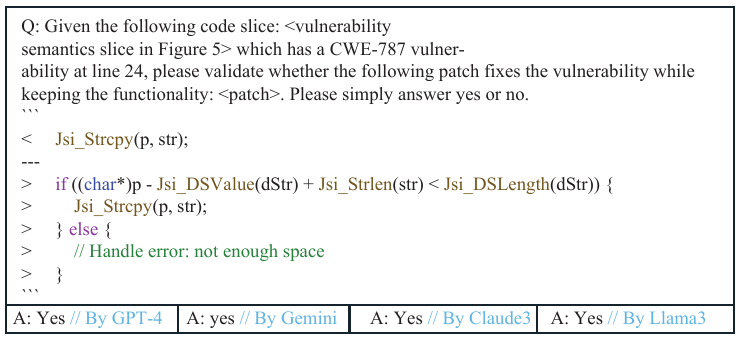}
        \vspace{-10pt}
        \caption{Multi-faceted patch validation for the sample in Figure~\ref{fig:example}.}
	\label{fig:discussion-valid}
        \vspace{-10pt}
\end{figure}

Algorithm~\ref{algo:validation} shows the workflow for the validation. It initializes by considering all candidate patches as potentially valid (line 2). It then enters a nested loop where each candidate patch is evaluated independently against each LLM to ascertain its validity (lines 3-4). Within the inner loop, each patch undergoes a validation check by an individual LLM, which determines if the patch is effective (line 5).
Each LLM evaluates whether each generated patch fixes the vulnerability correctly with the following prompt:
\vspace{-6pt}
\begin{tcolorbox}[left=1pt,right=1pt,top=0pt, bottom=3pt, lifted shadow={1mm}{1mm}{1mm}{1mm}{black!100!white}, colback=white, arc=0pt,auto outer arc, boxrule=.5pt]
    {Q: Given the following code slice: $\mathcal{V}_{cause}$ which has a vulnerability among $<$CWE-IDs$>$ and lines: $S_v$. Please validate whether the following patch fixes the vulnerability while keeping the functionality: $<$patch$>$. Please simply answer yes or no.}
\vspace{-4pt}
\end{tcolorbox}
\vspace{-4pt}

In the prompt, we also ask the LLMs whether the patch keeps the functionality to avoid valid but incorrect patches. 

Figure~\ref{fig:discussion-valid} shows the multi-faceted patch validation prompts and answers for the example in Figure~\ref{fig:example}. If any LLM approves the patch, a flag (retain) is set to True, indicating that the patch should be retained in the list of valid patches (lines 6-8).
If after evaluation by all the LLMs, a patch is not validated by any (i.e., the retain flag remains False), it is removed from the list of valid patches (lines 9-10). This step ensures that only those patches that have been validated by at least one LLM are kept.
The algorithm concludes by returning the refined list of validated patches, which have passed through this rigorous validation process (line 11). 
We use such an approach rather than majority voting because {\tech} aims to improve the recall while keeping the precision, which is more practically valuable than otherwise for real-world developers. Majority voting may improve the precision but reduce the recall, which is less helpful for real-world patching.

\vspace{-0pt}
\section{Implementation}
\vspace{-0pt}


To support the generation of \textit{interprocedural} vulnerability semantics, 
we use Joern~\cite{joern} to construct the SDG of the given vulnerable program under patching. 
In this work, since we evaluate {\tech} on a number of C code samples without compiling them, we chose Joern which is a powerful tool designed for scalable code analysis without requiring compilability of the input code.
Specifically, we first obtain the PDG (which includes the control and data dependencies) for each function (which Joern immediately provides). Then, we leverage the call graph also provided by Joern to conduct the interprocedural dependence analysis hence building the SDG. 
With the SDG, we extract the interprocedural backward slice from the vulnerable statements. In this work, we reuse the script from SySeVR~\cite{li2021sysevr} to quickly query the control flow nodes for the respective statements and find the in-degree edges and nodes. Based on this, we further implement the semantics-aware scoping algorithm, as shown in Algorithm~\ref{algo:semanticsscoped}. The extracted vulnerability semantics slices are stored as sliced source code text for LLMs to use.

\vspace{-0pt}
\section{Evaluation}
\vspace{-0pt}

We evaluate {\tech} via 
four research questions (RQs): 

\begin{itemize}[leftmargin=*,itemsep=0pt, topsep=0pt]
\item
\textbf{RQ1:} 
How effective is {\tech} for real-world vulnerabilities patching?
\item
\textbf{RQ2:} 
How does each of the key design components of {\tech} contribute to its overall performance?
\item
\textbf{RQ3:} 
How effective is {\tech} compared to existing vulnerability patching techniques?
\item
\textbf{RQ4:} 
How efficient is {\tech} for patch generation?
    
\end{itemize}

\noindent
We first describe the LLMs, datasets and metrics used. Then, we answer the four research questions above.

\vspace{2pt}\noindent
\textbf{LLMs.}
\begin{table}[tp]
    \vspace{-6pt}
    \centering
     \caption{LLMs used in our study}
    \scalebox{0.55}{
    \begin{tabular}{c|c|c|c|c|c|c}
    \hline
    Model & Version & \#Params& Max \#Tokens & Vendor & Release date & Cutoff Date \\ \hline
    Gemini-1.5 & gemini-1.5-pro & 50T & 30,720 & Google & May 2024 & Nov. 2023\\
    Claude-3.5 & claude-3.5-sonnet & 70B & 200,000 & Anthropic & Jun. 2024 & Apr. 2024 \\
    GPT-4 & gpt-4-turbo & 175B+ & 128,000 & OpenAI & May 2024 & Oct. 2023\\ 
    Llama-3.1 & llama-3.1-70b & 70B & 4,096 & Meta & July 2024 & Dec. 2023\\
    \hline
    \end{tabular}}
    \vspace{-5pt}
    \label{tab:llm}
\end{table}
Table~\ref{tab:llm} summarizes the four SOTA LLMs chosen for our study to cover the latest models of a variety of sizes and vendors. \textbf{Gemini-1.5}~\cite{team2023gemini} and \textbf{Claude-3.5}~\cite{claude3} are the latest LLMs from Google and Anthropic, respectively. 
We chose their most popular powerful ({\tt Pro} and {\tt Sonnet})
for both effectiveness and efficiency. \textbf{GPT-4} is the latest and most powerful LLM from OpenAI and we use its most powerful version {\tt GPT-4-Turbo}.
\textbf{Llama3.1}~\cite{touvron2023llama} is an SOTA LLM from Meta, of which we also chose to use the powerful version llama-3.1-70b for our experiments. We chose general-purpose LLMs instead of code-specific ones (e.g., CodeLlama) because the latter are less capable of code analysis. As shown in Appendix Table~\ref{tab:detailed-codellm}, our preliminary experiments revealed that {\tech}, when based on popular code-specific LLMs (CodeLlama~\cite{codellama}, CodeQwen 1.5~\cite{codeqwen}, and DeepSeek-Coder-v2~\cite{zhu2024deepseek}), achieved only 1.21\%, 10.12\%, and 9.09\% F1 scores, respectively, significantly lower than the performance of general-purpose LLMs when used in {\tech} as base LLMs as shown in Table~\ref{tab:patch-acc}.

\noindent
\textbf{Datasets.} To collect exemplar samples for Phase 1 (\textit{Exemplar Mining}), we use existing widely used vulnerability datasets {\tt \textbf{PatchDB}}~\cite{wang2021patchdb} and {\tt \textbf{CVEFixes}}~\cite{bhandari2021cvefixes}, where PatchDB contains 12073 real-world fixing samples and CVEFixes contains 4,120 real-world fixing samples in C language. In this work, we select the samples with the most popular CWEs in C languages: CWE-787 (out-of-bound write), CWE-125 (out-of-bound read), CWE-190 (integer overflow), CWE-401 (memory leak), CWE-457 (use of uninitialized variable), and CWE-476 (use of NULL pointer). However, some of the fixes contain edits not relevant to vulnerabilities (e.g., code refactoring or adding new functionalities). Therefore, we manually inspect each of the samples and filter out the inappropriate samples. To be realistic, we read the vulnerability reports and label vulnerability manifestation locations, rather than the code changed lines in the vulnerability fixing commits. We applied an inter-rater agreement/consensus procedure---each author independently labeled them, followed by cross-checking outcomes and discussions to resolve disagreements. We also remove the overlapped samples between PatchDB and CVEFixes, as well as confirming that there is no overlap between the training set and the testing sets discussed below. As a result, we collect 306 vulnerability fixing samples. 

For the testing set, considering that the latest powerful LLMs are trained on existing data, it is possible that they have seen the samples in PatchDB and CVEFixes during the training. Thus, there may be data leakage and contamination issues if we use the samples from PatchDB and CVEFixes as the testing set. Therefore, we collect a {\tt \textbf{Zero-Day dataset}}
where all the vulnerabilities were reported after the latest cutoff date of the LLMs (04/2024 from Claude-3.5-Sonnet). As a result, we collect 97 vulnerabilities which also cover the same CWEs as PatchDB and CVEFixes. Among them, 21 vulnerabilities are interprocedural where the vulnerability manifestation locations and the patching locations are in different functions. These samples cover 18 open-source projects including Linux Kernal, FFmpeg, etc. Appendix Table~\ref{tab:detailed-datasets} shows the detailed statistics of the dataset.

To compare {\tech} with existing vulnerability patching techniques, we also use an existing vulnerability dataset {\tt \textbf{ExtractFix}}~\cite{gao2021beyond} for comparison experiments. The samples from ExtractFix are compilable and come with test cases that allow us to validate patches. From this dataset, we collect 20 vulnerabilities that can be reproduced, for each of which we label the manifestation locations and the CWE IDs based on the crash information from the test cases. Detailed statistics of this dataset can be also found in Appendix Table~\ref{tab:detailed-datasets}. We also addressed data leakage concerns with it in Appendix~\ref{append:leakage}.

\vspace{2pt}\noindent
\textbf{Metrics.} Since the dataset covers many projects and it is difficult to compile each of the samples and use the test cases to validate the generated patches (furthermore, some of the samples do not come with test cases), we cannot automatically validate the generated patches if they do not exactly match the ground truths. Therefore, we manually check each generated patch. To improve the manual evaluation and better understand the generated patches, we consider four metrics: (1) if the generated patch exactly matches the ground truth, we consider it as \emph{syntactic equivalent (SynEq)}; (2) if the generated patch does not exactly match the ground truth but the behavior is the same as the ground truth, we consider it as \emph{semantically equivalent (SemEq)}; (3) if the patch has different behavior from the ground truth but still fix the vulnerability without breaking the code functionality, we consider it as \emph{plausible}; (4) all the samples with the three metrics above are considered as \emph{correct}.
To better simulate the real-world patching scenarios, we prompt the LLMs to generate up to five patches so that developers can pick up the one most suitable to patch the vulnerable code. To better evaluate the generated patches, we consider three measures: 
\begin{equation}
    recall=\frac{\#fixed\ samples}{\#testing\ sample}
\end{equation}
\begin{equation}
    precision=\frac{\#correct\ patches}{\#generated\ patches}
\end{equation}
\begin{equation}
    F1=\frac{2\times recall \times precision}{recall+precision}
\end{equation}
where $fixed\ sample$ means a sample where at least one of the generated patches fix the vulnerability.

\begin{table*}[tp]
  \centering
     \caption{Effectiveness of {\tech}}
     \vspace{2pt}
  \scalebox{0.43}{
\begin{tabular}{|c|l|r|r|r|r|r|r|r|r|r|r|r|r|r|r|r|r|r|r|r|r|r|r|r|r|}
\hline
\multirow{3}[6]{*}{Model} & \multicolumn{1}{c|}{\multirow{3}[6]{*}{Approach}} & \multicolumn{12}{c|}{Zero-Day Dataset}                                                        & \multicolumn{12}{c|}{ExtractFix Dataset} \\
\cline{3-26}      &       & \multicolumn{3}{c|}{\textbf{SynEq}} & \multicolumn{3}{c|}{\textbf{SemEq}} & \multicolumn{3}{c|}{\textbf{Plausible}} & \multicolumn{3}{c|}{\textbf{Correct}} & \multicolumn{3}{c|}{\textbf{SynEq}} & \multicolumn{3}{c|}{\textbf{SemEq}} & \multicolumn{3}{c|}{\textbf{Plausible}} & \multicolumn{3}{c|}{\textbf{Correct}} \\
\cline{3-26}      &       & \multicolumn{1}{l|}{\textbf{Recall}} & \multicolumn{1}{l|}{\textbf{Prec}} & \multicolumn{1}{l|}{\textbf{F1}} & \multicolumn{1}{l|}{\textbf{Recall}} & \multicolumn{1}{l|}{\textbf{Prec}} & \multicolumn{1}{l|}{\textbf{F1}} & \multicolumn{1}{l|}{\textbf{Recall}} & \multicolumn{1}{l|}{\textbf{Prec}} & \multicolumn{1}{l|}{\textbf{F1}} & \multicolumn{1}{l|}{\textbf{Recall}} & \multicolumn{1}{l|}{\textbf{Prec}} & \multicolumn{1}{l|}{\textbf{F1}} & \multicolumn{1}{l|}{\textbf{Recall}} & \multicolumn{1}{l|}{\textbf{Prec}} & \multicolumn{1}{l|}{\textbf{F1}} & \multicolumn{1}{l|}{\textbf{Recall}} & \multicolumn{1}{l|}{\textbf{Prec}} & \multicolumn{1}{l|}{\textbf{F1}} & \multicolumn{1}{l|}{\textbf{Recall}} & \multicolumn{1}{l|}{\textbf{Prec}} & \multicolumn{1}{l|}{\textbf{F1}} & \multicolumn{1}{l|}{\textbf{Recall}} & \multicolumn{1}{l|}{\textbf{Prec}} & \multicolumn{1}{l|}{\textbf{F1}} \\
\hline
\multirow{8}[2]{*}{GPT-4} & {\tech}  & \cellcolor[rgb]{ .776,  .878,  .706}6.19\% & \cellcolor[rgb]{ .776,  .878,  .706}1.64\% & \cellcolor[rgb]{ .776,  .878,  .706}\textit{\textbf{2.59\%}} & \cellcolor[rgb]{ .776,  .878,  .706}26.80\% & \cellcolor[rgb]{ .776,  .878,  .706}14.48\% & \cellcolor[rgb]{ .776,  .878,  .706}\textit{\textbf{18.80\%}} & \cellcolor[rgb]{ .776,  .878,  .706}25.77\% & \cellcolor[rgb]{ .776,  .878,  .706}12.84\% & \cellcolor[rgb]{ .776,  .878,  .706}\textit{\textbf{17.14\%}} & \cellcolor[rgb]{ .776,  .878,  .706}39.18\% & \cellcolor[rgb]{ .776,  .878,  .706}28.96\% & \cellcolor[rgb]{ .776,  .878,  .706}\textit{\textbf{33.30\%}} & \cellcolor[rgb]{ .776,  .878,  .706}5.00\% & \cellcolor[rgb]{ .776,  .878,  .706}2.30\% & \cellcolor[rgb]{ .776,  .878,  .706}\textit{\textbf{3.15\%}} & \cellcolor[rgb]{ .776,  .878,  .706}75.00\% & \cellcolor[rgb]{ .776,  .878,  .706}42.53\% & \cellcolor[rgb]{ .776,  .878,  .706}\textit{\textbf{54.28\%}} & \cellcolor[rgb]{ .776,  .878,  .706}40.00\% & \cellcolor[rgb]{ .776,  .878,  .706}10.34\% & \cellcolor[rgb]{ .776,  .878,  .706}\textit{\textbf{16.44\%}} & \cellcolor[rgb]{ .776,  .878,  .706}90.00\% & \cellcolor[rgb]{ .776,  .878,  .706}55.17\% & \cellcolor[rgb]{ .776,  .878,  .706}\textit{\textbf{68.41\%}} \\
      & No Validation & \cellcolor[rgb]{ .886,  .937,  .855}7.22\% & \cellcolor[rgb]{ .886,  .937,  .855}1.46\% & \cellcolor[rgb]{ .886,  .937,  .855}\textit{2.42\%} & \cellcolor[rgb]{ .886,  .937,  .855}28.87\% & \cellcolor[rgb]{ .886,  .937,  .855}13.10\% & \cellcolor[rgb]{ .886,  .937,  .855}\textit{18.02\%} & \cellcolor[rgb]{ .886,  .937,  .855}28.87\% & \cellcolor[rgb]{ .886,  .937,  .855}11.02\% & \cellcolor[rgb]{ .886,  .937,  .855}\textit{15.95\%} & \cellcolor[rgb]{ .886,  .937,  .855}44.33\% & \cellcolor[rgb]{ .886,  .937,  .855}25.57\% & \cellcolor[rgb]{ .886,  .937,  .855}\textit{32.43\%} & \cellcolor[rgb]{ .886,  .937,  .855}5.00\% & \cellcolor[rgb]{ .886,  .937,  .855}1.98\% & \cellcolor[rgb]{ .886,  .937,  .855}\textit{2.84\%} & \cellcolor[rgb]{ .886,  .937,  .855}80.00\% & \cellcolor[rgb]{ .886,  .937,  .855}41.58\% & \cellcolor[rgb]{ .886,  .937,  .855}\textit{54.72\%} & \cellcolor[rgb]{ .886,  .937,  .855}45.00\% & \cellcolor[rgb]{ .886,  .937,  .855}9.90\% & \cellcolor[rgb]{ .886,  .937,  .855}\textit{16.23\%} & \cellcolor[rgb]{ .886,  .937,  .855}90.00\% & \cellcolor[rgb]{ .886,  .937,  .855}53.47\% & \cellcolor[rgb]{ .886,  .937,  .855}\textit{67.08\%} \\
      & No Slicing & 3.09\% & 0.61\% & \textit{1.02\%} & 28.87\% & 8.98\% & \textit{13.70\%} & 21.65\% & 9.80\% & \textit{13.49\%} & 39.18\% & 19.39\% & \textit{25.94\%} & 10.00\% & 1.98\% & \textit{3.31\%} & 65.00\% & 41.58\% & \textit{50.72\%} & 30.00\% & 11.88\% & \textit{17.02\%} & 80.00\% & 55.45\% & \textcolor[rgb]{ 1,  0,  0}{\textit{65.50\%}} \\
      & Random Exemplars & 5.15\% & 1.02\% & \textit{1.71\%} & 31.96\% & 11.48\% & \textit{16.89\%} & 23.71\% & 6.97\% & \textit{10.77\%} & 44.33\% & 19.47\% & \textcolor[rgb]{ 1,  0,  0}{\textit{27.05\%}} & 10.00\% & 1.98\% & \textit{3.31\%} & 75.00\% & 35.64\% & \textit{48.32\%} & 50.00\% & 17.82\% & \textit{26.28\%} & 80.00\% & 55.45\% & \textit{65.50\%} \\
      & Manual Exemplars & 3.09\% & 1.03\% & \textit{1.54\%} & 28.87\% & 11.11\% & \textit{16.05\%} & 17.53\% & 5.97\% & \textit{8.90\%} & 42.27\% & 18.11\% & \textit{25.35\%} & 5.00\% & 0.99\% & \textit{1.65\%} & 70.00\% & 39.60\% & \textit{50.59\%} & 35.00\% & 11.88\% & \textit{17.74\%} & 85.00\% & 52.48\% & \textit{64.89\%} \\
      & Direct Reasoning & 3.09\% & 0.62\% & \textit{1.03\%} & 22.68\% & 8.28\% & \textit{12.13\%} & 19.59\% & 6.83\% & \textit{10.13\%} & 36.08\% & 15.73\% & \textit{21.91\%} & 5.00\% & 0.99\% & \textit{1.65\%} & 75.00\% & 43.56\% & \textit{55.11\%} & 30.00\% & 6.93\% & \textit{11.26\%} & 80.00\% & 51.49\% & \textit{62.65\%} \\
      & Standard Prompting & 3.09\% & 0.62\% & 1.03\% & 18.56\% & 7.19\% & 10.36\% & 17.53\% & 6.78\% & 9.77\% & 30.93\% & 14.58\% & 19.82\% & 5.00\% & 0.99\% & 1.65\% & 65.00\% & 40.59\% & 49.98\% & 30.00\% & 9.90\% & 14.89\% & 65.00\% & 51.49\% & 57.46\% \\
      & Zero-shot Completion & 0.00\% & 0.00\% & 0.00\% & 0.00\% & 0.00\% & 0.00\% & 12.37\% & 6.78\% & 8.76\% & 12.37\% & 6.78\% & 8.76\% & 0.00\% & 0.00\% & 0.00\% & 0.00\% & 0.00\% & 0.00\% & 15.00\% & 3.96\% & 6.27\% & 15.00\% & 3.96\% & 6.27\% \\
\hline
\multirow{8}[2]{*}{Gemini-1.5} & {\tech}  & \cellcolor[rgb]{ .776,  .878,  .706}2.06\% & \cellcolor[rgb]{ .776,  .878,  .706}0.56\% & \cellcolor[rgb]{ .776,  .878,  .706}\textit{\textbf{0.88\%}} & \cellcolor[rgb]{ .776,  .878,  .706}9.28\% & \cellcolor[rgb]{ .776,  .878,  .706}6.16\% & \cellcolor[rgb]{ .776,  .878,  .706}\textit{\textbf{7.41\%}} & \cellcolor[rgb]{ .776,  .878,  .706}21.64\% & \cellcolor[rgb]{ .776,  .878,  .706}9.80\% & \cellcolor[rgb]{ .776,  .878,  .706}\textit{\textbf{13.49\%}} & \cellcolor[rgb]{ .776,  .878,  .706}26.80\% & \cellcolor[rgb]{ .776,  .878,  .706}16.53\% & \cellcolor[rgb]{ .776,  .878,  .706}\textit{\textbf{20.44\%}} & \cellcolor[rgb]{ .776,  .878,  .706}0.00\% & \cellcolor[rgb]{ .776,  .878,  .706}0.00\% & \cellcolor[rgb]{ .776,  .878,  .706}\textit{\textbf{0.00\%}} & \cellcolor[rgb]{ .776,  .878,  .706}35.00\% & \cellcolor[rgb]{ .776,  .878,  .706}18.68\% & \cellcolor[rgb]{ .776,  .878,  .706}\textit{\textbf{24.20\%}} & \cellcolor[rgb]{ .776,  .878,  .706}30.00\% & \cellcolor[rgb]{ .776,  .878,  .706}16.48\% & \cellcolor[rgb]{ .776,  .878,  .706}\textit{\textbf{21.28\%}} & \cellcolor[rgb]{ .776,  .878,  .706}50.00\% & \cellcolor[rgb]{ .776,  .878,  .706}35.16\% & \cellcolor[rgb]{ .776,  .878,  .706}\textit{\textbf{42.90\%}} \\
      & No Validation & \cellcolor[rgb]{ .886,  .937,  .855}2.06\% & \cellcolor[rgb]{ .886,  .937,  .855}0.55\% & \cellcolor[rgb]{ .886,  .937,  .855}\textit{0.87\%} & \cellcolor[rgb]{ .886,  .937,  .855}9.28\% & \cellcolor[rgb]{ .886,  .937,  .855}6.32\% & \cellcolor[rgb]{ .886,  .937,  .855}\textit{7.52\%} & \cellcolor[rgb]{ .886,  .937,  .855}21.64\% & \cellcolor[rgb]{ .886,  .937,  .855}9.61\% & \cellcolor[rgb]{ .886,  .937,  .855}\textit{13.31\%} & \cellcolor[rgb]{ .886,  .937,  .855}26.80\% & \cellcolor[rgb]{ .886,  .937,  .855}16.48\% & \cellcolor[rgb]{ .886,  .937,  .855}\textit{20.41\%} & \cellcolor[rgb]{ .886,  .937,  .855}0.00\% & \cellcolor[rgb]{ .886,  .937,  .855}0.00\% & \cellcolor[rgb]{ .886,  .937,  .855}\textit{0.00\%} & \cellcolor[rgb]{ .886,  .937,  .855}35.00\% & \cellcolor[rgb]{ .886,  .937,  .855}17.46\% & \cellcolor[rgb]{ .886,  .937,  .855}\textit{23.30\%} & \cellcolor[rgb]{ .886,  .937,  .855}30.00\% & \cellcolor[rgb]{ .886,  .937,  .855}15.87\% & \cellcolor[rgb]{ .886,  .937,  .855}\textit{20.76\%} & \cellcolor[rgb]{ .886,  .937,  .855}50.00\% & \cellcolor[rgb]{ .886,  .937,  .855}33.33\% & \cellcolor[rgb]{ .886,  .937,  .855}\textit{40.00\%} \\
      & No Slicing & 0.00\% & 0.00\% & \textit{0.00\%} & 10.31\% & 5.08\% & \textit{6.81\%} & 18.56\% & 10.56\% & \textit{13.69\%} & 24.74\% & 15.93\% & \textcolor[rgb]{ 1,  0,  0}{\textit{19.38\%}} & 0.00\% & 0.00\% & \textit{0.00\%} & 20.00\% & 15.94\% & \textit{17.74\%} & 35.00\% & 18.84\% & \textit{24.50\%} & 40.00\% & 34.78\% & \textit{37.21\%} \\
      & Random Exemplars & 1.03\% & 0.34\% & \textit{0.51\%} & 12.37\% & 5.82\% & \textit{7.92\%} & 15.46\% & 8.56\% & \textit{11.02\%} & 24.74\% & 14.72\% & \textit{18.46\%} & 5.00\% & 1.09\% & \textit{1.79\%} & 35.00\% & 17.39\% & \textit{23.24\%} & 25.00\% & 10.87\% & \textit{15.15\%} & 45.00\% & 29.35\% & \textit{35.53\%} \\
      & Manual Exemplars & 2.06\% & 0.68\% & \textit{1.03\%} & 9.28\% & 4.11\% & \textit{5.70\%} & 17.52\% & 10.96\% & \textit{13.48\%} & 22.68\% & 15.75\% & \textit{18.59\%} & 0.00\% & 0.00\% & \textit{0.00\%} & 25.00\% & 13.79\% & \textit{17.78\%} & 30.00\% & 14.94\% & \textit{19.95\%} & 45.00\% & 28.74\% & \textit{35.07\%} \\
      & Direct Reasoning & 1.03\% & 0.29\% & 0.46\% & 8.25\% & 3.24\% & 4.66\% & 20.61\% & 8.85\% & 12.39\% & 25.77\% & 12.38\% & 16.73\% & 0.00\% & 0.00\% & 0.00\% & 45.00\% & 30.21\% & 36.15\% & 10.00\% & 3.13\% & 4.76\% & 45.00\% & 33.33\% & \textcolor[rgb]{ 1,  0,  0}{38.30\%} \\
      & Standard Prompting & 1.03\% & 0.26\% & 0.41\% & 11.34\% & 4.10\% & 6.03\% & 18.56\% & 6.92\% & 10.08\% & 25.77\% & 11.28\% & 15.69\% & 0.00\% & 0.00\% & 0.00\% & 20.00\% & 12.87\% & 15.66\% & 30.00\% & 12.87\% & 18.01\% & 45.00\% & 25.74\% & 32.75\% \\
      & Zero-shot Completion & 0.00\% & 0.00\% & 0.00\% & 0.00\% & 0.00\% & 0.00\% & 3.09\% & 1.21\% & 1.74\% & 3.09\% & 1.21\% & 1.74\% & 0.00\% & 0.00\% & 0.00\% & 0.00\% & 0.00\% & 0.00\% & 20.00\% & 8.24\% & 11.67\% & 20.00\% & 8.24\% & 11.67\% \\
\hline
\multirow{8}[2]{*}{Claude-3.5} & {\tech}  & \cellcolor[rgb]{ .776,  .878,  .706}4.12\% & \cellcolor[rgb]{ .776,  .878,  .706}0.92\% & \cellcolor[rgb]{ .776,  .878,  .706}\textit{\textbf{1.51\%}} & \cellcolor[rgb]{ .776,  .878,  .706}37.11\% & \cellcolor[rgb]{ .776,  .878,  .706}17.32\% & \cellcolor[rgb]{ .776,  .878,  .706}\textit{\textbf{23.62\%}} & \cellcolor[rgb]{ .776,  .878,  .706}22.68\% & \cellcolor[rgb]{ .776,  .878,  .706}10.62\% & \cellcolor[rgb]{ .776,  .878,  .706}\textit{\textbf{14.47\%}} & \cellcolor[rgb]{ .776,  .878,  .706}49.48\% & \cellcolor[rgb]{ .776,  .878,  .706}28.87\% & \cellcolor[rgb]{ .776,  .878,  .706}\textit{\textbf{36.46\%}} & \cellcolor[rgb]{ .776,  .878,  .706}10.00\% & \cellcolor[rgb]{ .776,  .878,  .706}2.04\% & \cellcolor[rgb]{ .776,  .878,  .706}\textit{\textbf{3.39\%}} & \cellcolor[rgb]{ .776,  .878,  .706}80.00\% & \cellcolor[rgb]{ .776,  .878,  .706}44.90\% & \cellcolor[rgb]{ .776,  .878,  .706}\textit{\textbf{57.52\%}} & \cellcolor[rgb]{ .776,  .878,  .706}45.00\% & \cellcolor[rgb]{ .776,  .878,  .706}18.37\% & \cellcolor[rgb]{ .776,  .878,  .706}\textit{\textbf{26.09\%}} & \cellcolor[rgb]{ .776,  .878,  .706}85.00\% & \cellcolor[rgb]{ .776,  .878,  .706}65.31\% & \cellcolor[rgb]{ .776,  .878,  .706}\textit{\textbf{73.86\%}} \\
      & No Validation & \cellcolor[rgb]{ .886,  .937,  .855}5.15\% & \cellcolor[rgb]{ .886,  .937,  .855}1.07\% & \cellcolor[rgb]{ .886,  .937,  .855}\textit{1.78\%} & \cellcolor[rgb]{ .886,  .937,  .855}37.11\% & \cellcolor[rgb]{ .886,  .937,  .855}16.52\% & \cellcolor[rgb]{ .886,  .937,  .855}\textit{22.87\%} & \cellcolor[rgb]{ .886,  .937,  .855}22.68\% & \cellcolor[rgb]{ .886,  .937,  .855}10.09\% & \cellcolor[rgb]{ .886,  .937,  .855}\textit{13.96\%} & \cellcolor[rgb]{ .886,  .937,  .855}50.52\% & \cellcolor[rgb]{ .886,  .937,  .855}27.68\% & \cellcolor[rgb]{ .886,  .937,  .855}\textit{35.77\%} & \cellcolor[rgb]{ .886,  .937,  .855}10.00\% & \cellcolor[rgb]{ .886,  .937,  .855}1.98\% & \cellcolor[rgb]{ .886,  .937,  .855}\textit{3.31\%} & \cellcolor[rgb]{ .886,  .937,  .855}80.00\% & \cellcolor[rgb]{ .886,  .937,  .855}44.55\% & \cellcolor[rgb]{ .886,  .937,  .855}\textit{57.23\%} & \cellcolor[rgb]{ .886,  .937,  .855}45.00\% & \cellcolor[rgb]{ .886,  .937,  .855}17.82\% & \cellcolor[rgb]{ .886,  .937,  .855}\textit{25.53\%} & \cellcolor[rgb]{ .886,  .937,  .855}85.00\% & \cellcolor[rgb]{ .886,  .937,  .855}64.36\% & \cellcolor[rgb]{ .886,  .937,  .855}\textit{73.25\%} \\
      & No Slicing & 3.09\% & 0.63\% & \textit{1.05\%} & 27.84\% & 12.47\% & \textit{17.23\%} & 27.84\% & 10.57\% & \textit{15.32\%} & 41.24\% & 23.68\% & \textcolor[rgb]{ 1,  0,  0}{\textit{30.08\%}} & 10.00\% & 1.98\% & \textit{3.31\%} & 65.00\% & 44.55\% & \textit{52.87\%} & 30.00\% & 7.92\% & \textit{12.53\%} & 80.00\% & 54.46\% & \textit{64.80\%} \\
      & Random Exemplars & 1.03\% & 0.21\% & \textit{0.34\%} & 28.87\% & 15.11\% & \textit{19.84\%} & 17.53\% & 6.83\% & \textit{9.83\%} & 41.24\% & 22.15\% & \textit{28.82\%} & 5.00\% & 0.99\% & \textit{1.65\%} & 75.00\% & 43.56\% & \textit{55.11\%} & 45.00\% & 16.83\% & \textit{24.50\%} & 90.00\% & 61.39\% & \textcolor[rgb]{ 1,  0,  0}{\textit{72.99\%}} \\
      & Manual Exemplars & 2.06\% & 0.42\% & \textit{0.69\%} & 29.90\% & 13.13\% & \textit{18.24\%} & 21.65\% & 8.13\% & \textit{11.82\%} & 41.24\% & 21.67\% & \textit{28.41\%} & 5.00\% & 0.99\% & \textit{1.65\%} & 80.00\% & 46.53\% & \textit{58.84\%} & 40.00\% & 9.90\% & \textit{15.87\%} & 85.00\% & 57.43\% & \textit{68.54\%} \\
      & Direct Reasoning & 2.06\% & 0.49\% & 0.80\% & 23.71\% & 14.32\% & 17.86\% & 17.53\% & 7.65\% & 10.66\% & 37.11\% & 22.47\% & 27.99\% & 5.00\% & 0.99\% & 1.65\% & 75.00\% & 52.48\% & 61.75\% & 35.00\% & 15.84\% & 21.81\% & 75.00\% & 69.31\% & 72.04\% \\
      & Standard Prompting & 2.06\% & 0.43\% & 0.72\% & 24.74\% & 12.61\% & 16.70\% & 14.43\% & 6.09\% & 8.56\% & 28.87\% & 19.13\% & 23.01\% & 15.00\% & 2.97\% & 4.96\% & 55.00\% & 40.59\% & 46.71\% & 25.00\% & 12.87\% & 16.99\% & 65.00\% & 56.44\% & 60.42\% \\
      & Zero-shot Completion & 0.00\% & 0.00\% & 0.00\% & 4.12\% & 1.12\% & 1.76\% & 15.46\% & 8.28\% & 10.78\% & 18.56\% & 9.40\% & 12.48\% & 0.00\% & 0.00\% & 0.00\% & 5.00\% & 3.49\% & 4.11\% & 30.00\% & 17.44\% & 22.06\% & 30.00\% & 20.93\% & 24.66\% \\
\hline
\multirow{8}[2]{*}{Llama-3.1} & {\tech}  & \cellcolor[rgb]{ .776,  .878,  .706}1.03\% & \cellcolor[rgb]{ .776,  .878,  .706}0.27\% & \cellcolor[rgb]{ .776,  .878,  .706}\textit{\textbf{0.42\%}} & \cellcolor[rgb]{ .776,  .878,  .706}21.65\% & \cellcolor[rgb]{ .776,  .878,  .706}9.55\% & \cellcolor[rgb]{ .776,  .878,  .706}\textit{\textbf{13.25\%}} & \cellcolor[rgb]{ .776,  .878,  .706}18.56\% & \cellcolor[rgb]{ .776,  .878,  .706}8.75\% & \cellcolor[rgb]{ .776,  .878,  .706}\textit{\textbf{11.89\%}} & \cellcolor[rgb]{ .776,  .878,  .706}35.05\% & \cellcolor[rgb]{ .776,  .878,  .706}18.57\% & \cellcolor[rgb]{ .776,  .878,  .706}\textit{\textbf{24.28\%}} & \cellcolor[rgb]{ .776,  .878,  .706}5.00\% & \cellcolor[rgb]{ .776,  .878,  .706}1.03\% & \cellcolor[rgb]{ .776,  .878,  .706}\textit{\textbf{1.71\%}} & \cellcolor[rgb]{ .776,  .878,  .706}70.00\% & \cellcolor[rgb]{ .776,  .878,  .706}44.33\% & \cellcolor[rgb]{ .776,  .878,  .706}\textit{\textbf{54.28\%}} & \cellcolor[rgb]{ .776,  .878,  .706}20.00\% & \cellcolor[rgb]{ .776,  .878,  .706}8.25\% & \cellcolor[rgb]{ .776,  .878,  .706}\textit{\textbf{11.68\%}} & \cellcolor[rgb]{ .776,  .878,  .706}80.00\% & \cellcolor[rgb]{ .776,  .878,  .706}53.61\% & \cellcolor[rgb]{ .776,  .878,  .706}\textit{\textbf{64.20\%}} \\
      & No Validation & \cellcolor[rgb]{ .886,  .937,  .855}1.03\% & \cellcolor[rgb]{ .886,  .937,  .855}0.22\% & \cellcolor[rgb]{ .886,  .937,  .855}\textit{0.36\%} & \cellcolor[rgb]{ .886,  .937,  .855}22.68\% & \cellcolor[rgb]{ .886,  .937,  .855}9.41\% & \cellcolor[rgb]{ .886,  .937,  .855}\textit{13.30\%} & \cellcolor[rgb]{ .886,  .937,  .855}20.62\% & \cellcolor[rgb]{ .886,  .937,  .855}8.09\% & \cellcolor[rgb]{ .886,  .937,  .855}\textit{11.62\%} & \cellcolor[rgb]{ .886,  .937,  .855}38.14\% & \cellcolor[rgb]{ .886,  .937,  .855}17.72\% & \cellcolor[rgb]{ .886,  .937,  .855}\textit{24.20\%} & \cellcolor[rgb]{ .886,  .937,  .855}5.00\% & \cellcolor[rgb]{ .886,  .937,  .855}0.99\% & \cellcolor[rgb]{ .886,  .937,  .855}\textit{1.65\%} & \cellcolor[rgb]{ .886,  .937,  .855}70.00\% & \cellcolor[rgb]{ .886,  .937,  .855}42.57\% & \cellcolor[rgb]{ .886,  .937,  .855}\textit{52.95\%} & \cellcolor[rgb]{ .886,  .937,  .855}20.00\% & \cellcolor[rgb]{ .886,  .937,  .855}7.92\% & \cellcolor[rgb]{ .886,  .937,  .855}\textit{11.35\%} & \cellcolor[rgb]{ .886,  .937,  .855}80.00\% & \cellcolor[rgb]{ .886,  .937,  .855}51.49\% & \cellcolor[rgb]{ .886,  .937,  .855}\textit{62.65\%} \\
      & No Slicing & 2.06\% & 0.49\% & \textit{0.79\%} & 20.62\% & 9.09\% & \textit{12.62\%} & 11.34\% & 4.67\% & \textit{6.61\%} & 28.87\% & 14.25\% & \textit{19.08\%} & 0.00\% & 0.00\% & \textit{0.00\%} & 70.00\% & 29.70\% & \textit{41.71\%} & 30.00\% & 10.89\% & \textit{15.98\%} & 75.00\% & 40.59\% & \textit{52.68\%} \\
      & Random Exemplars & 1.03\% & 0.22\% & \textit{0.37\%} & 15.46\% & 6.52\% & \textit{9.17\%} & 9.28\% & 4.49\% & \textit{6.06\%} & 23.71\% & 11.24\% & \textit{15.25\%} & 5.00\% & 0.99\% & \textit{1.65\%} & 65.00\% & 39.60\% & \textit{49.22\%} & 10.00\% & 2.97\% & \textit{4.58\%} & 75.00\% & 43.56\% & \textit{55.11\%} \\
      & Manual Exemplars & 3.09\% & 0.74\% & \textit{1.19\%} & 19.59\% & 11.30\% & \textit{14.33\%} & 11.34\% & 4.67\% & \textit{6.61\%} & 29.90\% & 16.71\% & \textcolor[rgb]{ 1,  0,  0}{\textit{21.44\%}} & 5.00\% & 0.99\% & \textit{1.65\%} & 70.00\% & 34.65\% & \textit{46.36\%} & 30.00\% & 11.88\% & \textit{17.02\%} & 85.00\% & 47.52\% & \textit{60.96\%} \\
      & Direct Reasoning & \textit{4.12\%} & \textit{0.83\%} & \textit{1.39\%} & \textit{15.46\%} & \textit{7.71\%} & \textit{10.29\%} & \textit{17.52\%} & \textit{6.45\%} & \textit{9.44\%} & \textit{31.95\%} & \textit{15.00\%} & \textit{20.41\%} & \textit{0.00\%} & \textit{0.00\%} & \textit{0.00\%} & \textit{70.00\%} & \textit{48.51\%} & \textit{57.31\%} & \textit{30.00\%} & \textit{5.94\%} & \textit{9.92\%} & \textit{70.00\%} & \textit{54.46\%} & \textcolor[rgb]{ 1,  0,  0}{\textit{61.26\%}} \\
      & Standard Prompting & \textit{3.09\%} & \textit{0.63\%} & \textit{1.05\%} & \textit{14.43\%} & \textit{7.34\%} & \textit{9.73\%} & \textit{14.43\%} & \textit{5.87\%} & \textit{8.35\%} & \textit{24.74\%} & \textit{13.84\%} & \textit{17.75\%} & \textit{0.00\%} & \textit{0.00\%} & \textit{0.00\%} & \textit{60.00\%} & \textit{40.59\%} & \textit{48.43\%} & \textit{15.00\%} & \textit{5.94\%} & \textit{8.51\%} & \textit{65.00\%} & \textit{46.53\%} & \textit{54.24\%} \\
      & Zero-shot Completion & \textit{0.00\%} & \textit{0.00\%} & \textit{0.00\%} & \textit{0.00\%} & \textit{0.00\%} & \textit{0.00\%} & \textit{9.28\%} & \textit{4.17\%} & \textit{5.75\%} & \textit{9.28\%} & \textit{4.17\%} & \textit{5.75\%} & \textit{0.00\%} & \textit{0.00\%} & \textit{0.00\%} & \textit{5.00\%} & \textit{1.98\%} & \textit{2.84\%} & \textit{5.00\%} & \textit{2.97\%} & \textit{3.73\%} & \textit{5.00\%} & \textit{4.95\%} & \textit{4.98\%} \\
\hline
\end{tabular}%
}
\label{tab:patch-acc}
\end{table*}

\vspace{-0pt}
\subsection{RQ1: Effectiveness}\label{ssec:rq1}
\vspace{-0pt}

\noindent
Table~\ref{tab:patch-acc} shows the effectiveness of {\tech} against the two testing datasets. We notice that {\tech} achieves the best overall effectiveness for vulnerability patching. With {\tech}, GPT-4, Gemini-1.5, Claude-3.5, and Llama-3.1 achieve 33.30\%, 20.44\%, 36.46\%, and 24.28\% F1 for the correctness metric on our collected Zero-Day dataset, which is the best among other prompting approaches. 
On the existing ExtractFix dataset, {\tech} again achieves the best among other prompting approaches, with 68.41\%, 42.90\%, 73.86\%, and 64.20\% F1 for the correctness metric, respectively. 
The overall effectiveness on ExtractFix dataset is higher, indicating the challenges on Zero-Day vulnerability patching. 

In terms of recall, Claude-3.5 achieves the best with 49.48\% on the Zero-Day dataset and GPT-4 acheives the best with 90.00\% on the ExtractFix dataset. 
GPT-4 achieves the best precision 28.96\% on the Zero-Day dataset and Claude-3.5 achieves the best precision 65.31\% on the ExtractFix dataset. 
In terms of different metrics, Claude-3.5 has 37.11\% and 80.00\% recall on semantics equivalent against the Zero-Day and ExtractFix datasets, respectively. This indicates that {\tech} generates patches close to developers' patches.

\vspace{-0pt}
\begin{tcolorbox}[left=1pt,right=1pt,top=0pt, bottom=2pt, lifted shadow={1mm}{-2mm}{3mm}{0.1mm}{black!50!white},arc=0pt,auto outer arc, boxrule=.5pt,leftrule=2pt]
{\tech} is effective for vulnerability patching, with up to 36.46\% and 73.86\% F1 on the Zero-Day and ExtractFix dataset, respectively, showing its practical potential.
\vspace{-4pt}
\end{tcolorbox}
\vspace{-0pt}

\begin{table}[h]
\vspace{-8pt}
  \centering
     \caption{Effectiveness of {\tech} against the interprocedural samples from the Zero-Day dataset}
  \scalebox{0.5}{

\begin{tabular}{|l|c|r|r|r|c|r|r|r|}
\hline
\multicolumn{1}{|c|}{\multirow{2}[0]{*}{\textbf{Approach}}} & \multirow{2}[0]{*}{\textbf{Model}} & \multicolumn{3}{c|}{\textbf{Correct}} & \multirow{2}[0]{*}{\textbf{Model}} & \multicolumn{3}{c|}{\textbf{Correct}} \\
\cline{3-5}\cline{7-9}      &       & \multicolumn{1}{l|}{\textbf{Recall}} & \multicolumn{1}{l|}{\textbf{Prec}} & \multicolumn{1}{l|}{\textbf{F1}} &       & \multicolumn{1}{l|}{\textbf{Recall}} & \multicolumn{1}{l|}{\textbf{Prec}} & \multicolumn{1}{l|}{\textbf{F1}} \\
\hline
{\tech}  & \multirow{8}[2]{*}{GPT-4} & \cellcolor[rgb]{ .776,  .878,  .706}38.10\% & \cellcolor[rgb]{ .776,  .878,  .706}26.67\% & \cellcolor[rgb]{ .776,  .878,  .706}\textit{\textbf{31.37\%}} & \multirow{8}[2]{*}{Claude-3.5} & \cellcolor[rgb]{ .776,  .878,  .706}47.62\% & \cellcolor[rgb]{ .776,  .878,  .706}25.93\% & \cellcolor[rgb]{ .776,  .878,  .706}\textit{\textbf{33.57\%}} \\
No Validation &       & \cellcolor[rgb]{ .886,  .937,  .855}38.10\% & \cellcolor[rgb]{ .886,  .937,  .855}25.23\% & \cellcolor[rgb]{ .886,  .937,  .855}\textit{30.35\%} &       & \cellcolor[rgb]{ .886,  .937,  .855}47.62\% & \cellcolor[rgb]{ .886,  .937,  .855}25.23\% & \cellcolor[rgb]{ .886,  .937,  .855}\textit{32.98\%} \\
No Slicing &       & 33.33\% & 18.18\% & \textit{23.53\%} &       & 38.10\% & 21.78\% & \textit{27.72\%} \\
Random Exemplars &       & 33.33\% & 15.32\% & \textit{20.99\%} &       & 47.62\% & 19.82\% & \textcolor[rgb]{ 1,  0,  0}{\textit{27.99\%}} \\
Manual Exemplars &       & 38.10\% & 17.27\% & \textcolor[rgb]{ 1,  0,  0}{\textit{23.77\%}} &       & 42.86\% & 18.92\% & 26.25\% \\
Direct Reasoning &       & 38.10\% & 13.51\% & \textit{19.95\%} &       & 42.86\% & 19.81\% & \textit{27.10\%} \\
Standard Prompting &       & 19.05\% & 6.60\% & 9.81\% &       & 14.29\% & 6.86\% & 9.27\% \\
Zero-shot Completion &       & 9.52\% & 4.72\% & 6.31\% &       & 14.29\% & 5.94\% & 8.39\% \\
\hline
{\tech}  & \multirow{8}[2]{*}{Gemini-1.5} & \cellcolor[rgb]{ .776,  .878,  .706}28.57\% & \cellcolor[rgb]{ .776,  .878,  .706}14.14\% & \cellcolor[rgb]{ .776,  .878,  .706}\textit{\textbf{18.91\%}} & \multirow{8}[2]{*}{Llama-3.1} & \cellcolor[rgb]{ .776,  .878,  .706}38.10\% & \cellcolor[rgb]{ .776,  .878,  .706}17.65\% & \cellcolor[rgb]{ .776,  .878,  .706}\textit{\textbf{24.12\%}} \\
No Validation &       & \cellcolor[rgb]{ .886,  .937,  .855}28.57\% & \cellcolor[rgb]{ .886,  .937,  .855}13.34\% & \cellcolor[rgb]{ .886,  .937,  .855}\textit{18.18\%} &       & \cellcolor[rgb]{ .886,  .937,  .855}38.10\% & \cellcolor[rgb]{ .886,  .937,  .855}14.41\% & \cellcolor[rgb]{ .886,  .937,  .855}\textit{20.91\%} \\
No Slicing &       & 23.81\% & 12.37\% & \textcolor[rgb]{ 1,  0,  0}{\textit{16.28\%}} &       & 14.29\% & 23.08\% & \textcolor[rgb]{ 1,  0,  0}{\textit{17.65\%}} \\
Random Exemplars &       & 23.81\% & 11.46\% & \textit{15.47\%} &       & 23.81\% & 10.89\% & \textit{14.95\%} \\
Manual Exemplars &       & 23.81\% & 11.11\% & 15.15\% &       & \textit{19.05\%} & \textit{15.38\%} & \textit{17.02\%} \\
Direct Reasoning &       & 23.81\% & 7.62\% & \textit{11.54\%} &       & 28.57\% & 10.81\% & \textit{15.68\%} \\
Standard Prompting &       & 4.76\% & 1.10\% & 1.79\% &       & \textit{9.52\%} & \textit{3.13\%} & \textit{4.71\%} \\
Zero-shot Completion &       & 0.00\% & 0.00\% & 0.00\% &       & \textit{9.52\%} & \textit{2.70\%} & \textit{4.21\%} \\
\hline
\end{tabular}

}
\vspace{-8pt}
\label{tab:interprocedural}
\end{table}

\vspace{-6pt}
\subsection{RQ2: Contributions of Components}
\vspace{-0pt}
As shown in Table~\ref{tab:patch-acc}, we then examine the contributions of the {\tech} components with the ablated prompting approaches. The numbers marked in {\textcolor{red}{red}} are the best F1 scores achieved by the ablated prompting approaches. 
One of the ablated versions, ``Manual Exemplars'', corresponds to our prior work~\cite{nong2024chain}, which is \textit{also an LLM-based 
patching baseline}. 

When we remove the multi-faceted patch validation, as shown in the "no validation", we notice that the precision drops and thus the F1 becomes lower, with 20.41\%--35.77\% F1 on the Zero-Day dataset and 40.00\%--73.25\% F1 on the ExtractFix dataset, indicating that the multi-faceted patch validation is useful to improve the precision. 


We then examine the contribution of semantics-aware scoping. In this ablation study, instead of using the vulnerability semantics slice, we directly input the whole source code sample into the LLMs for analysis. As shown in rows ``No slicing'', without the vulnerability semantics slicing, the recall, precision, and F1 drop significantly compared with {\tech}, with 19.08\%--30.08\% F1 on the Zero-Day dataset and 37.21\%--65.50\% F1 on the ExtractFix dataset. The reason is that, without slicing, the code not relevant to vulnerabilities is also inputted into the LLMs, thus distracting the analysis and impact the accuracy of the root cause analysis. To prove this, we examine the correct rates of the reasoning on the testing samples. As Table~\ref{tab:reasoning} shows, vulnerability semantics reasoning achieves higher accuracy if we have vulnerability semantics scoping/slicing. This indicates the importance of inputting core vulnerability semantics for analysis.

To examine the usefulness of dynamic adaptive prompting, we conduct experiments with randomly selected exemplars rather than selecting exemplars based on the root causes of the vulnerabilities. Note that we still keep the CWEs of the exemplars the same as the testing sample. As shown in the rows ``Random Exemplars'', we notice that the F1 scores drop dramatically compared with {\tech}.
The exemplars selected for prompting have promising fixing strategies which help LLMs patch the vulnerability correctly, while they may pick up incorrect fixing strategies to patch with random exemplars. This indicates the importance and effectiveness of dynamic prompting 
for vulnerability patching. 

As the exemplars in our exemplar pool are generated by LLMs, to examine whether these LLM-generated exemplars are better than human curated exemplars, we also manually write exemplars following the same reasoning steps. 
As shown in rows ``Manual Exemplars'', the overall effectiveness drops significantly, indicating the effectiveness of adaptive prompting for our LLM-based vulnerability patching design.

\begin{figure}[h]
\centering
\vspace{2pt}
	\includegraphics[width=0.9\linewidth]{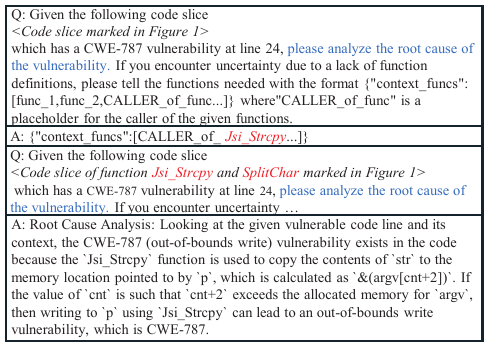}
        \vspace{-8pt}
        \caption{Root cause analysis generated by GPT-4 \emph{\textbf{without} vulnerability semantics reasoning} for the sample in Figure~\ref{fig:example}.}
	\label{fig:example-cause-nosem}
        \vspace{-0pt}
\end{figure}

To show the effectiveness of vulnerability semantics reasoning, we simply prompt the LLMs to generate root causes, as shown in Figure~\ref{fig:example-cause-nosem}. As shown in the row ``Direct Reasoning'', the overall effectiveness drops significantly compared to {\tech}. The reason is that the accuracy of root cause analysis is lower than vulnerability semantics reasoning, as shown in Table~\ref{tab:reasoning}. This indicates that guiding LLMs for vulnerability semantics reasoning is effective for improving vulnerability analysis and patching.

To show the patching effectiveness of the LLMs without any specific prompting design, we directly ask the LLMs to patch the given vulnerabilities. As shown in row ``Standard Prompting'', we notice that the standard prompting have much lower effectiveness compared with {\tech}, achieving only up to 23.01\% F1 on the Zero-Day dataset and 60.42\% F1 on the ExtractFix dataset even with the state-of-the-art LLMs. 

Recently, Pearce et al.\cite{pearce2023examining} approaches the LLM-based vulnerability repair problem through zero-shot code completion, which removes the vulnerable code in the program and lets the LLMs fill the vulnerable parts so as to fix the vulnerability. We use their best template {\tt s2} for comparison. As shown in the Row ``zero-shot completion'', this approach has the lowest effectiveness, with up to 12.48\% F1 on our Zero-Day dataset and 24.66\% F1 on the ExtractFix dataset, indicating that only providing prior code without further information is insufficient for LLMs to fix the vulnerabilities.




\vspace{-0pt}
\begin{tcolorbox}[left=1pt,right=1pt,top=0pt, bottom=2pt, lifted shadow={1mm}{-2mm}{3mm}{0.1mm}{black!50!white},arc=0pt,auto outer arc, boxrule=.5pt,leftrule=2pt]
Each component in {\tech} 
substantially contributes to its effectiveness, 
justifying its overall design. 
\vspace{-4pt}
\end{tcolorbox}
\vspace{-0pt}

We further evaluate {\tech}'s effectiveness on interprocedural vulnerabilities using the 21 interprocedural samples from the Zero-Day dataset. Table~\ref{tab:interprocedural} shows the correctness results (the complete results are in Appendix Table~\ref{tab:detailed-interprocedural}). Similarly, the numbers marked in {\textcolor{red}{red}} are the best F1 scores achieved by the ablated prompting approaches. 
Compared to the overall results on the complete Zero-Day dataset, {\tech} and its ablated versions show a modest decrease in performance (e.g., {\tech}'s F1 score with GPT-4 decreases from 33.30\% to 31.37\%). 

In contrast, the baseline approaches—Standard Prompting and Zero-Shot Completion—exhibit a substantial decline in effectiveness (e.g., Claude-3.5 with standard prompting and zero-shot completion decrease from 23.01\% and 12.48\% to 9.81\% and 6.31\%, respectively). Note that for Standard Prompting, we provide all the functions covered by the computed vulnerability semantics. For zero-shot completion, we directly provide functions with the ground-truth patch location as it can only work in such a setting. These already benefit the two baselines but they still fall short with these interprocedural samples. 
This deterioration can be attributed to the absence of interprocedural analysis in these approaches which adversely affects the LLMs' performance. 

\vspace{-0pt}
\begin{tcolorbox}[left=1pt,right=1pt,top=0pt, bottom=2pt, lifted shadow={1mm}{-2mm}{3mm}{0.1mm}{black!50!white},arc=0pt,auto outer arc, boxrule=.5pt,leftrule=2pt]
While {\tech} has a slight performance drop on exclusively interprocedural vulnerabilities, baseline approaches show a dramatic decline. This highlights the effectiveness merits of interprocedural analysis in {\tech}.
\vspace{-4pt}
\end{tcolorbox}
\vspace{-0pt}

\begin{table}[t]
  \centering
     \caption{Comparison to existing techniques in terms of recall}
  \scalebox{0.6}{
\begin{tabular}{|l|r|r|r|r|c|c|c|c|}
\hline
      & \multicolumn{4}{c|}{ExtractFix Dataset} & \multicolumn{4}{c|}{Zero-Day Dataset} \\
\hline
Technique & \multicolumn{1}{l|}{SynEq} & \multicolumn{1}{l|}{SemEq} & \multicolumn{1}{l|}{Plausible} & \multicolumn{1}{l|}{Correct} & \multicolumn{1}{l|}{SynEq} & \multicolumn{1}{l|}{SemEq} & \multicolumn{1}{l|}{Plausible} & \multicolumn{1}{l|}{Correct} \\
\hline
{\tech} & \cellcolor[rgb]{ .776,  .878,  .706}5\% & \cellcolor[rgb]{ .776,  .878,  .706}75\% & \cellcolor[rgb]{ .776,  .878,  .706}40\% & \cellcolor[rgb]{ .776,  .878,  .706}\textbf{90\%} & \multicolumn{1}{r|}{\cellcolor[rgb]{ .776,  .878,  .706}4.12\%} & \multicolumn{1}{r|}{\cellcolor[rgb]{ .776,  .878,  .706}37.11\%} & \multicolumn{1}{r|}{\cellcolor[rgb]{ .776,  .878,  .706}22.68\%} & \multicolumn{1}{r|}{\cellcolor[rgb]{ .776,  .878,  .706}\textbf{49.48\%}} \\
VulRepair & 10\%  & 25\%  & 0\%   & 35\%  & \multicolumn{1}{r|}{2.06\%} & \multicolumn{1}{r|}{15.46\%} & \multicolumn{1}{r|}{1.03\%} & \multicolumn{1}{r|}{\textcolor[rgb]{ 1,  0,  0}{17.53\%}} \\
Getafix & 20\%  & 5\%   & 0\%   & 25\%  & \multicolumn{1}{r|}{5.15\%} & \multicolumn{1}{r|}{0.00\%} & \multicolumn{1}{r|}{1.03\%} & \multicolumn{1}{r|}{6.19\%} \\
ExtractFix & 20\%  & 35\%  & 30\%  & \textcolor[rgb]{ 1,  0,  0}{85\%} & -     & -     & -     & - \\
VulnFix & 20\%  & 10\%  & 15\%  & 45\%  & -     & -     & -     & - \\
\hline
\end{tabular}%
}
\label{tab:comparison}
\vspace{-6pt}
\end{table}

\vspace{-0pt}
\subsection{RQ3: Comparison to Existing Techniques}
\vspace{-0pt}
To compare the effectiveness of {\tech} to existing vulnerability patching techniques, we select four state-of-the-art non-LLM-based tools which cover different technical categories, as shown in Table~\ref{tab:comparison}, where the {\textcolor{red}{red}} numbers indicate the best recall achieved by the baselines. VulRepair~\cite{fu2022vulrepair} is a DL-based vulnerability repair technique which fine-tunes a pre-trained programming language model CodeT5~\cite{wang2021codet5} to repair vulnerabilities. 
Getafix~\cite{bader2019getafix} is a data-driven pattern-based vulnerability repair technique which learns vulnerability patching patterns from existing samples. 
ExtractFix~\cite{gao2021beyond} is a traditional vulnerability patching technique which leverages test cases and constraints to fix the vulnerabilities. VulnFix~\cite{zhang2022program} is a SOTA vulnerability patching technique based on inductive inference. 

As ExtractFix and VulnFix require compilable projects and test cases which are difficult to build on our Zero-Day dataset, we first use the ExtractFix dataset for comparison. As ExtractFix and VulnFix generate many patches for testing, we only compare the recall metric in this RQ. For VulRepair and Getafix which are learning-based, we use our PatchDB+CVEFixes training set to re-train their models.

As shown in Table~\ref{tab:comparison}, on the ExtractFix dataset, {\tech} with the best LLM (GPT-4) achieves the best recall (90\%) compared with the baseline techniques. Note that the recall sum of syntactic equivalent, semantically equivalent, and plausible may be over recall on correct because there are multiple patches generated for each testing sample. ExtractFix achieves 85\% recall because it leverages the test cases and constraints for patching. Without such information, other baseline techniques only achieve 25\%-45\% recall. 

Note that the ExtractFix dataset is released in 2021, which is possible to cause the data leakage issue as it is before the cutoff date of the used LLMs. Thus, we test the LLMs whether they know the related information and patches given the meta data. The results show that they either do not know or give wrong answers, indicating that the data leakage issue is not serious. Detailed results can be found in Appendix~\ref{append:leakage}. 

We further evaluate the baselines on the Zero-Day dataset. Since these samples are not compilable, we limit our comparison to VulRepair and Getafix as the other two (VulnFix and ExtractFix) both require that the input code is compilable. As shown in Table~\ref{tab:comparison}, {\tech} significantly outperforms both baselines, achieving a recall of 49.48\% compared to VulRepair's 17.53\% and Getafix's 6.19\%. This highlights {\tech}'s superior effectiveness in patching.

\vspace{-0pt}
\begin{tcolorbox}[left=1pt,right=1pt,top=0pt, bottom=3pt, lifted shadow={1mm}{-2mm}{3mm}{0.1mm}{black!50!white},arc=0pt,auto outer arc, boxrule=.5pt,leftrule=2pt]
Compared to traditional techniques, {\tech} is effective in vulnerability patching without extra information.
\vspace{-4pt}
\end{tcolorbox}
\vspace{-0pt}

\vspace{-0pt}
\subsection{RQ4: Efficiency}
\vspace{-0pt}
We evaluate the efficiency of {\tech} in terms of time cost and token usage. Table~\ref{tab:efficiency} shows the time cost and token usage of {\tech} per sample on each LLM. After mining the exemplars, each sample takes between 37.148 seconds and 50.209 seconds to generate a patch, including the time for selecting the exemplars and generating the candidate patches. In this process, the input/context token usage ranges from 5,684 to 6,802 per sample, and the generated token usage ranges from 584 to 886. We notice that Claude-3.5 and GPT-4 take more tokens while Gemini-1.5 and Llama-3.1 take less. The reason is that the former two models tend to generate more tokens when generating the exemplars and patches.
Based on the rate of each model's API, patching each sample costs between 0.2 and 8 cents. However, Gemini-1.5 provides a free version if the usage is not large and Llama-3.1 supports deploying locally. If so, the two models are free.

\vspace{-0pt}
\begin{tcolorbox}[left=1pt,right=1pt,top=0pt, bottom=2pt, lifted shadow={1mm}{-2mm}{3mm}{0.1mm}{black!50!white},arc=0pt,auto outer arc, boxrule=.5pt,leftrule=2pt]
{\tech} is efficient for vulnerability patching in terms of both time and money cost.
\vspace{-4pt}
\end{tcolorbox}
\vspace{-0pt}

\begin{table}
 \centering
     \caption{Time and token cost of {\tech} per sample} 
\vspace{4pt}
\scalebox{0.65}{
\begin{tabular}{|c|c|c|c|c|}
    \hline
    \textbf{Model} & \textbf{Time cost} & \textbf{Contexts Tokens} & \textbf{Generated Tokens} & \textbf{Price}  \\ \hline
    GPT-4 & 40.169s & 6290 & 647 & \$0.0823 \\ 
    Claude-3.5 & 50.209s & 6802 & 886 & \$0.0336 \\ 
    Gemini-1.5 & 37.148s & 5684 & 584 & \$0.0026 \\ 
    Llama-3.1 & 46.598s & 5749 & 591 & \$0.0203 \\ \hline
\end{tabular}
\vspace{-4pt}
}
\label{tab:efficiency}
\end{table}

\vspace{-0pt}
\section{Discussion}
\vspace{-0pt}
We further discuss the advantages of {\tech} that enable its greater effectiveness, as well as why and how it fails on some samples, its tool usability, its current limitations, and the extensibility of {\tech}. 
\vspace{-0pt}
\subsection{Why {\tech} Works}
There are several advantages of {\tech} for vulnerability patching. The first advantage is contributed by the semantics-aware scoping. As Figure~\ref{fig:example} shows, without semantics-aware scoping, the code to be analyzed and patched includes many statements not relevant to the vulnerabilities. In this case, the LLMs are distracted and thus are more difficult to generate correct root cause and patches. In comparison, resulting from our semantics-aware scoping, the slice analyzed and patched by LLMs only includes the core code fragment for vulnerability analysis and patching. This not only helps LLMs concentrate on the vulnerability semantics which is the most important parts for patching, but also makes the data and control flow clearer than the original code. For example, in the original code, the variable {\tt cnt} at line 22 is 6 lines away from its last definition at line 16. However, in the vulnerability semantics slice, it is only two lines away, which makes LLMs easier to find the data dependencies for vulnerability patching.

Figures~\ref{fig:discussion-std}, \ref{fig:discussion-s2} and \ref{fig:discussion-appatch} show the generated analysis and patch by GPT-4 with standard prompting, zero-shot completion and {\tech}, where the standard prompting and zero-shot completion fail to patch the vulnerability correctly while {\tech} succeeds. We notice that in {\tech} with semantics-aware scoping, GPT-4 is able to capture more data flow information for variables {\tt p}, {\tt cnt}, and {\tt dStr}, while the GPT-4 with standard prompting and zero-shot code completion only analyze and patch the code based on the vulnerable line itself. This indicates that semantics-aware scoping helps LLMs better work on the data dependency analysis, which is important for effective vulnerability patching.

The second advantage of {\tech} is that it prompts dynamically based on the testing samples rather than using random or fixed prompts. This allows LLMs to leverage a large amount of existing real-world vulnerability data with few-shot learning, while avoiding LLM fine-tuning which incurs high costs of time and money.  
Because of the token limitation and the disadvantage of LLMs for long-text process, it is not feasible to input all the exemplars for few-shot learning. In contrast, the dynamic adaptive prompting with CoT guides LLMs to analyze the vulnerabilities in depth rather than analyzing them superficially. As Figures~\ref{fig:discussion-std}, \ref{fig:discussion-s2} and \ref{fig:discussion-appatch} show, GPT-4 with {\tech} conducts in-depth analysis for the root cause of the vulnerability, while GPT-4 with standard prompting and zero-shot completion cannot. The exemplars from dynamic adaptive prompting guide GPT-4 to find the size and boundary of the pointer {\tt p}, which significantly helps generate the patch correctly.

Notably, the ability of LLMs to automatically generate the vulnerability semantics reasoning steps (i.e., the natural language representation of vulnerability semantics) is key to the automation in Step 1.2. Also, since the reasoning essentially dissects the vulnerable behavior of the code, they immediately serve as root cause analysis. Importantly, using \textit{functionality semantics} instead would help the LLMs understand the functional behaviors of the code, hence not helping much the models with understanding the vulnerability root cause hence guiding their patch generation, as we extensively validated. 


The third advantage is that {\tech} conducts multi-faceted patch validation strategy to remove the unusable patches. As shown in Table~\ref{tab:patch-acc}, with multi-facted patch validation, the precision is improved significantly without impacting the recall. This improves the F1 and effectively reduces the load of developers for selecting appropriate patches, thus shows the practicability of {\tech}.

\begin{table}
 \centering
     \caption{Accuracy of CodeQL detection, localization, classification, and the three combined against the Zero-Day dataset} 
\vspace{4pt}
\scalebox{0.65}{
\begin{tabular}{|l|r|r|r|r|}
\hline
\textbf{Scenario} & \multicolumn{1}{l|}{\textbf{Detection}} & \multicolumn{1}{l|}{\textbf{Localization}} & \multicolumn{1}{l|}{\textbf{Classification}} & \multicolumn{1}{l|}{\textbf{Combined}} \\
\hline
Fully Automated & 87.62\% & 59.79\% & 69.07\% & 55.67\% \\
Realistic & 87.62\% & 72.16\% & 83.50\% & 68.04\% \\
\hline
\end{tabular}
}
\vspace{-4pt}
\label{tab:codeql}
\end{table}

\begin{table}
 \centering
     \caption{End-to-end results 
     against the Zero-Day dataset}
\vspace{4pt}
\scalebox{0.6}{
\begin{tabular}{|c|c|c|r|r|r|}
\hline
\multirow{2}[0]{*}{\textbf{Scenario}} & \multirow{2}[0]{*}{\textbf{Model}} & \multirow{2}[0]{*}{\textbf{Approach}} & \multicolumn{3}{c|}{\textbf{Correct}} \\
\cline{4-6}      &       &       & \multicolumn{1}{l|}{\textbf{Recall}} & \multicolumn{1}{l|}{\textbf{Prec}} & \multicolumn{1}{l|}{\textbf{F1}} \\
\hline
\multirow{6}[0]{*}{Fully Automated} & \multirow{4}[2]{*}{Claude-3.5 } & \multicolumn{1}{l|}{{\tech}} & \cellcolor[rgb]{ .776,  .878,  .706}34.02\% & \cellcolor[rgb]{ .776,  .878,  .706}18.36\% & \cellcolor[rgb]{ .776,  .878,  .706}\textit{\textbf{23.85\%}} \\
      &       & \multicolumn{1}{l|}{Manual Exemplars} & 22.68\% & 11.33\% & \textit{15.11\%} \\
      &       & \multicolumn{1}{l|}{Standard Prompting} & 16.49\% & 11.15\% & 13.30\% \\
      &       & \multicolumn{1}{l|}{Zero-shot Completion} & 15.46\% & 9.04\% & 11.41\% \\
\cline{2-6}      & VulRepair & -     & 16.49\% & 9.18\% & 11.79\% \\
\cline{2-6}      & Getafix & -     & 6.19\% & 5.70\% & 5.93\% \\
\hline
\multirow{6}[0]{*}{Realistic} & \multirow{4}[2]{*}{Claude-3.5 } & \multicolumn{1}{l|}{{\tech}} & \cellcolor[rgb]{ .776,  .878,  .706}41.24\% & \cellcolor[rgb]{ .776,  .878,  .706}20.40\% & \cellcolor[rgb]{ .776,  .878,  .706}\textit{\textbf{27.29\%}} \\
      &       & \multicolumn{1}{l|}{Manual Exemplars} & 25.77\% & 11.38\% & \textit{15.79\%} \\
      &       & \multicolumn{1}{l|}{Standard Prompting} & 24.74\% & 16.16\% & 19.55\% \\
      &       & \multicolumn{1}{l|}{Zero-shot Completion} & 15.46\% & 9.04\% & 11.41\% \\
\cline{2-6}      & VulRepair & -     & 16.49\% & 9.18\% & 11.79\% \\
\cline{2-6}      & Getafix & -     & 6.19\% & 5.70\% & 5.93\% \\
\hline
\end{tabular}
}
\label{tab:usability}
\end{table}

\vspace{-0pt}
\subsection{Usability of {\tech}}\label{ssec:usability}
\vspace{-0pt}
In this work, we assume that the vulnerabilities under patching are pre-detected with correct CWE and location information. 
However, such information may not be always available. In this case, {\tech} may need to collaborate with other vulnerability analyzers (that provide the information) to work in practice. Thus, to demonstrate that our formulation of {\tech} is \emph{realistic} and showcase its usability, we evaluate its performance when fed with such information produced from upstream detection/localization and classification tools, a setting referred to as \textit{end-to-end} integration, against our Zero-Day dataset on the best-performing LLM Claude-3.5. 

We consider two integration scenarios for a comprehensive evaluation. 
The first is a \ul{fully automated} workflow in which both information from the upstream tools is directly fed to {\tech} as its inputs without human intervention.  
The second is a scenario in which the outputs of those tools are first inspected, verified, and calibrated if necessary, and the validated/calibrated location and CWE information is then fed to {\tech}. In a typical, realistic vulnerability management workflow, developers would inspect and calibrate the outputs of automated (vulnerability localization and classification) tools before proceeding with patching using those outputs, for many reasons (e.g., accuracy/trust issues---since no automated tools are perfect and they can produce false positives and suffer false negatives, among other considerations~\cite{flawfinderfool}). Thus, we refer to the second as the \ul{realistic scenario}. 

For the upstream tooling, we chose CodeQL~\cite{codeql} as it is an industry-grade security analyzer widely used on real-world projects. Also, as it serves detection/localization and classification purposes at the same time, users need just one tool instead of multiple to provide both inputs to {\tech}. 
For CodeQL to work specifically for a given project, users often customize the query packs based on the project information~\cite{codeql}. 
We also followed this common practice for the projects evaluated. 
Notably, our query-pack customization did not assume any knowledge about the ground-truth results (i.e., detection/localization/classification outcomes), avoiding biases and keeping the realism of the end-to-end integration setting. 
We extended the memory manipulation function set in the query packs so that the project’s custom memory-management APIs (e.g., {\tt kmalloc}, {\tt kfree}) can be scanned as well---which any ordinary user would be able to do as well for a given project to scan. 
For the realistic scenario involving developer inspection, we had a graduate student with 3-year relevant experience simulate the realistic human-intervention step between vulnerability localization/classification and patching, who inspected and calibrated CodeQL outputs.

{\tech}'s end-to-end performance results in both scenarios are in Table~\ref{tab:usability} (with further details in Appendix Table~\ref{tab:detailed-codeql}). 
To contextualize these results, Table~\ref{tab:codeql} gives the upstream tool performance, which is largely consistent with what earlier studies found on CodeQL~\cite{charoenwet2024empirical}. 
Note that CodeQL may report multiple locations and CWEs for one sample, similar to other vulnerability localization techniques~\cite{fu2022linevul,hin2022linevd,mirskyvulchecker}. Therefore, we follow their evaluation approach and report top-10 accuracy for localization and classification where the ranking is based on the security severity~\cite{codeql2}. 
As shown, {\tech} worked reasonably well in these end-to-end settings, with expected accuracy drops 
versus when perfect vulnerability location and CWE information is used (Table~\ref{tab:patch-acc}). 
These gaps can be justified by inaccuracies of the upstream results with/without human inspection. 
{\tech} also outperforms the baseline approaches, demonstrating its merits and superior practical performance. 
With more advanced upstream tools to provide more accurate vulnerability locations and CWEs, and users with more expertise (e.g., developers of the project under patching), these performance could be 
even better.

\subsection{How and Why {\tech} Fails} \label{sec:fail}
We conduct case studies on the all the failure cases of {\tech} on the Zero-Day dataset to investigate the symptoms of them. There are several major symptom groups. 

\begin{itemize}[leftmargin=*,itemsep=0pt, topsep=0pt]
    \item \textbf{Incorrect Vulnerability Identification.} The most common symptom is incorrect vulnerability identification, with 35.69\% of the symptoms. This category includes cases where the initial analysis misidentified the type or nature of the vulnerability. For example, as shown in Figure~\ref{fig:failure-1} the actual issue is related to a missing condition check for the debug event code type, not an out-of-bounds write as initially assumed. 
    \item 
    \textbf{Insufficient or incorrect code modification.} Another common issue was insufficient or incorrect code modification (31.23\% of symptoms). For example, as shown in Figure~\ref{fig:failure-2}, the patch focuses on local buffer handling while the actual fix require dynamic memory allocation and more comprehensive input validation. 
\end{itemize}

These symptoms show \textit{how} {\tech} fails to patch some samples. 
We further identified three main root causes: 

\begin{itemize}[leftmargin=*,itemsep=0pt, topsep=0pt]
    \item \textbf{Misunderstanding the actual vulnerabilities.} The most common root cause for failed patching is that the LLMs misunderstand the actual vulnerabilities, with 41.33\% of the root causes. This usually causes incorrect vulnerability identification, as shown in Figure~\ref{fig:failure-1}.
    \item \textbf{Inadequate analysis of code context and dependencies.} Another common root cause for failed patching is inadequate analysis of code context and dependencies (25.21\% of root causes). This usually causes insufficient or incorrect code modification, as shown in Figure~\ref{fig:failure-2}. 
    \item \textbf{Failure to consider all edge cases or scenarios.} This category covers situations where the patch didn't account for all possible scenarios or edge cases, with 10.21\% of the root causes. For example, in one sample, the LLM incorrectly handles special characters in shell quoting and fails to address all cases of characters requiring special treatment, particularly those needing quoting rather than escaping.
\end{itemize}



\subsection{Limitations}
There are several factors that may limit {\tech} in practice. The first limitation is that the root cause analysis may not be accurate at the testing phase. If the vulnerability root cause analysis is incorrect, the direction of patching is misled and the generated patches are less likely to be correct. 
The second limitation is that we only collect 306+76+20 patching samples for evaluation, because collecting these samples require manual works to label. Yet we still cannot ensure these samples are compilable and come with test cases. 

For the first limitation, {\tech} mitigates it by using the state-of-the-art and most powerful LLMs to generate the patches. We exclude the LLMs that are not capable of conducting logical analysis, such as Falcon~\cite{zxhang2023falcon}, StarCoder~\cite{li2023starcoder}, and GPT-NeoX~\cite{black2022gpt}. All the selected LLMs in this work have the capability of analyzing the code logically. 
For the second limitation, we collected as many as samples as we can so that they can cover more CWEs, vulnerable code patterns, and patching strategies/patterns.

Indeed, the two limitations can be attributed to the vulnerability datasets we use. In an ideal vulnerability patching scenario, the exploits should be provided and the dynamic tracing of the vulnerability should be available. However, the used datasets, PatchDB and CVEFixes, only provide the raw information of the vulnerabilities (e.g., versions, patches, source code) without further information. Considering that reproducing the vulnerabilities based on the CVE reports is very time-consuming, it is difficult for us to build a dataset with hundreds of compilable and vulnerability-reproducable code samples in a short time. At the same time, existing vulnerability datasets coming with compilable code and test cases are relatively small, such as the ExtractFix dataset~\cite{gao2021beyond} which only has 30 samples. Therefore, future works should try to build larger compilable vulnerability datasets so that rigorous static and dynamic analysis can be feasible. 

\vspace{-0pt}
\subsection{Extensibility}
\vspace{-0pt}
A common concern for the practicability of {\tech} is whether it can be extended to other CWEs and languages. While the current implementation and the evaluation are based on C language with six common CWEs, the design of {\tech} is language/CWE agnostic and is extendable to other CWEs and languages. To extend {\tech} to other CWEs, users can add exemplar samples of other CWEs to generate the respective exemplars. The samples with more CWEs can be also found in the PatchDB and CVEFixes datasets. To extend {\tech} to other languages, users can set up a code analyzer that is able to construct SDGs and then follow Algorithm~\ref{algo:semanticsscoped} to extract the vulnerability semantics slices. Meanwhile, our used code analyzer, Joern, also supports more than ten languages, including the most popular languages such as Java, Python, C\#, and Go.

\vspace{-0pt}
\subsection{LLMs for Automated Patching}
\vspace{-0pt}
Beyond the merits of {\tech} itself, 
our work also reveals several insights for future automated patching with LLMs.

The first is that prompting design matters for effective vulnerability patching. Based on our evaluation, simply asking LLMs to patch vulnerabilities work poorly because of the lack of guidance. While vulnerability patching by 
the SOTA approach~\cite{pearce2023examining} 
shows its potential when LLMs started surging, it performs poorly even with the latest and most powerful LLMs. This indicates that the prompts for vulnerability patching need to be specifically designed with comprehensive guidance.

The second insight is that LLM-based vulnerability patching still needs  traditional code analysis to complement. The main reason is that LLMs usually have token limitation so that we cannot input a large amount of code into the model. Even if some of the models support more tokens, the long input and output text may also distract the model from concentrating the important parts for patching because of the limitation of their base structure-Transformer~\cite{vaswani2017attention}. Therefore, extracting the important code slices by traditional code analysis techniques is promising for LLM-based vulnerability patching. 

The third insight is that dynamically prompting LLMs adaptively is necessary for few-shot learning-based vulnerability patching. The reason is that, because of the token limitation of LLMs, it is difficult to input all the samples covering different vulnerability root causes and patching strategies. The exemplars used for few-shot learning should be based on testing samples to achieve better effectiveness.

Meanwhile, we choose to use LLMs as they are, without pre-training and fine-tuning. The main reason is that the cost of LLMs also matters. If we choose to pre-train our own LLMs, the training data, hardware usage, and time cost would be considerable. Only big companies such as OpenAI and Google have enough resource to pre-train LLMs.
Currently, most of the powerful LLMs are commercial, thus it is difficult to access these LLMs directly for fine-tuning. Although some provide APIs for fine-tuning, the models are usually old (e.g., GPT-3.5) and the price is expensive considering the amount of data for fine-tuning. Therefore, prompting LLMs 
would be more feasible than pre-training and fine-tuning.


\section{Related Work}
Recent research has 
focused on example-based approaches to vulnerability detection and repair, demonstrating the effectiveness of learning from past vulnerabilities and their corresponding fixes.
Ma et al.~\cite{ma2017vurle} presented VuRLE, a 
system that utilizes machine learning to automatically detect and repair vulnerabilities by learning from examples. Similarly, Zhang et al.~\cite{zhang2022example} 
developed a similar approach for Java. 


Deep learning techniques have also shown significant promise in automating the vulnerability detection and repair process. Chen et al.~\cite{chen2022neural} explored neural transfer learning for repairing vulnerabilities in C code. VulRepair, a T5-based automated software vulnerability repair tool, is presented by Fu et al.~\cite{fu2022vulrepair}. This approach shows significant improvements in the accuracy and efficiency of vulnerability repair.

Program analysis and language models offer powerful techniques for understanding and fixing software vulnerabilities by analyzing code semantics and employing complicated models.
Gao et al.~\cite{gao2021beyond} proposed a novel method that goes beyond conventional testing by extracting crash constraints to guide program vulnerability repair.
Lastly, Pearce et al.~\cite{pearce2023examining} examined the application of LLM for zero-shot vulnerability repair. Their study highlights the potential of LLMs to fix vulnerabilities without extensive task-specific training, leveraging the models' ability to generalize from a vast amount of data.

\vspace{-6pt}
\section{Conclusion}
\vspace{-0pt}
We propose {\tech}, an automated vulnerability patching framework that features dynamic adaptive prompting on LLMs to elicit their effective reasoning for vulnerability root-cause analysis hence quality patch generation.  
We have demonstrated, on four latest LLMs, that {\tech} substantially outperforms both existing prompting approaches and state-of-the-art non-LLM-based patching techniques. 

\section*{Acknowledgments}
We thank our shepherd and the anonymous reviewers for their effective guidance and constructive comments. 
For this work, Yu Nong, Haoran Yang, and Haipeng Cai were supported by Army Research Office (ARO) under Grant No. W911NF-21-1-002, National Science Foundation (NSF) under Grant No. CCF-2146233 and CCF-2505223, and Office of Naval Research (ONR) under Grant No. N000142212111; 
Long Cheng was supported by NSF under Grant No. 2239605 and 2228616; and  
Hongxin Hu was supported by NSF under Grant No. 2228617, 2120369, and 2129164.


\section*{Ethics Considerations}
\vspace{-0pt}
Our research is focused on the development and evaluation of methods for patching software vulnerabilities, which complies with the ethical guidelines outlined by the USENIX Security'25 conference. Specifically, we have ensured that our study adheres to principles concerning the responsible handling of security research.

Our study focuses on generating and testing patches for vulnerabilities in software systems. 
While we evaluate our technique mainly in a zero-day setting (i.e., using the vulnerabilities that are zero-days to the LLMs which have a cutoff date before the disclosure time of these vulnerabilities), 
the vulnerabilities have already been disclosed (as CVEs) publicly by the time of our paper writing. 
Thus, our research itself does not cause harm or risks related to vulnerability disclosure, nor raising any new security threats. 
Nevertheless, throughout our research process, we have taken care to avoid any actions that could result in unintended harm, such as exposing systems or users to new security risks. We have ensured that all vulnerability data utilized is anonymized and does not contain any identifiable information about individuals or organizations. Besides, we have not conducted any experiments on production systems that could impact real-world users, hence avoiding any potential disruption or harm.

\vspace{-0pt}
\section*{Open Science} 
\vspace{-0pt}
We fully support the new open science policy of USENIX Security'25, which requires research results to be publicly accessible or for researchers to provide a valid explanation if this is not possible. To align with this policy, we commit to openly sharing all artifacts generated from our research, including the codebase, datasets, and experimental results, through a publicly accessible repository. 
The source code and documentation of {\tech} as well as our experimental results have been made 
available at 
\href{https://zenodo.org/records/14741018}{\underline{https://zenodo.org/records/14741018}}. 

In summary, our work aligns with both the ethical guidelines and open science policy of USENIX Security'25. We are committed to conducting research that not only adheres to high ethical standards but also promotes openness and community engagement in advancing cybersecurity research.

{
 \small
\bibliography{paper-abbr}

\begin{thebibliography}{10}

\bibitem{codeql}
{CodeQL} documentation.
\newblock \url{https://codeql.github.com/docs/}, 2021.

\bibitem{codellama}
{Introducing Code Llama}, {A} state-of-the-art large language model for coding.
\newblock \url{https://ai.meta.com/blog/code-llama-large-language-model-coding/}, 2023.

\bibitem{codeqwen}
Code with {CodeQwen1.5}.
\newblock \url{https://qwenlm.github.io/blog/codeqwen1.5/}, 2024.

\bibitem{flawfinderfool}
A fool with a tool is still a fool.
\newblock \url{https://dwheeler.com/flawfinder/}, 2024.

\bibitem{claude3}
Meet claude.
\newblock \url{https://www.anthropic.com/claude}, 2024.

\bibitem{codeql2}
Metadata for {CodeQL} queries.
\newblock \url{https://codeql.github.com/docs/writing-codeql-queries/metadata-for-codeql-queries/}, 2024.

\bibitem{bader2019getafix}
Johannes Bader, Andrew Scott, Michael Pradel, and Satish Chandra.
\newblock Getafix: Learning to fix bugs automatically.
\newblock {\em OOPSLA}, 3:1--27, 2019.

\bibitem{beszedes2002union}
{\'A}rp{\'a}d Besz{\'e}des, Csaba Farag{\'o}, Z~Mihaly Szabo, J{\'a}nos Csirik, and Tibor Gyim{\'o}thy.
\newblock Union slices for program maintenance.
\newblock In {\em ICSM}, pages 12--21, 2002.

\bibitem{bhandari2021cvefixes}
Guru Bhandari, Amara Naseer, and Leon Moonen.
\newblock {CVEfixes}: {Automated} collection of vulnerabilities and their fixes from open-source software.
\newblock In {\em PROMISE}, pages 30--39, 2021.

\bibitem{binkley2007empirical}
David Binkley, Nicolas Gold, and Mark Harman.
\newblock An empirical study of static program slice size.
\newblock {\em TOSEM}, 16(2):8--es, 2007.

\bibitem{black2022gpt}
Sidney Black, Stella Biderman, Eric Hallahan, Quentin Anthony, Leo Gao, Laurence Golding, Horace He, Connor Leahy, Kyle McDonell, Jason Phang, et~al.
\newblock {GPT-NeoX-20B}: An open-source autoregressive language model.
\newblock In {\em Proceedings of BigScience Episode\# 5--Workshop on Challenges \& Perspectives in Creating Large Language Models}, pages 95--136, 2022.

\bibitem{cai2023generating}
Haipeng Cai, Yu~Nong, Yuzhe Ou, and Feng Chen.
\newblock Generating vulnerable code via learning-based program transformations.
\newblock In {\em AI Embedded Assurance for Cyber Systems}, pages 123--138. Springer, 2023.

\bibitem{charoenwet2024empirical}
Wachiraphan Charoenwet, Patanamon Thongtanunam, Van-Thuan Pham, and Christoph Treude.
\newblock An empirical study of static analysis tools for secure code review.
\newblock In {\em ISSTA}, pages 691--703, 2024.

\bibitem{chen2022neural}
Zimin Chen, Steve Kommrusch, and Martin Monperrus.
\newblock Neural transfer learning for repairing security vulnerabilities in {C} code.
\newblock {\em TSE}, 49(1):147--165, 2022.

\bibitem{vulconsequence231}
{Ericsson}.
\newblock Software vulnerability: Impact \& ways to avoid it.
\newblock \url{https://www.ericsson.com/en/security/vulnerability-management}, 2023.

\bibitem{vulconsequence232}
{Forbes Technology Council}.
\newblock Zero-day vulnerabilities: 17 consequences and complications.
\newblock \url{https://www.forbes.com/councils/forbestechcouncil/2023/05/26/zero-day-vulnerabilities-17-consequences-and-complications}, 2023.

\bibitem{fu2022linevul}
Michael Fu and Chakkrit Tantithamthavorn.
\newblock {LineVul}: {A} {Transformer-based} line-level vulnerability prediction.
\newblock In {\em MSR}, pages 608--620, 2022.

\bibitem{fu2022vulrepair}
Michael Fu, Chakkrit Tantithamthavorn, Trung Le, Van Nguyen, and Dinh Phung.
\newblock {VulRepair}: {A T5-based} automated software vulnerability repair.
\newblock In {\em ESEC/FSE}, pages 935--947, 2022.

\bibitem{fu2021flowdist}
Xiaoqin Fu and Haipeng Cai.
\newblock {FlowDist}: Multi-staged refinement-based dynamic information flow analysis for distributed software systems.
\newblock In {\em USENIX Security Symposium}, pages 2093--2110, 2021.

\bibitem{gao2021beyond}
Xiang Gao, Bo~Wang, Gregory~J Duck, Ruyi Ji, Yingfei Xiong, and Abhik Roychoudhury.
\newblock Beyond tests: Program vulnerability repair via crash constraint extraction.
\newblock {\em TOSEM}, 30(2):1--27, 2021.

\bibitem{he2023controlling}
Jingxuan He and Martin Vechev.
\newblock Large language models for code: Security hardening and adversarial testing.
\newblock In {\em CCS}, pages 1865--1879, 2023.

\bibitem{hin2022linevd}
David Hin, Andrey Kan, Huaming Chen, and M~Ali Babar.
\newblock {LineVD}: {Statement-level} vulnerability detection using graph neural networks.
\newblock In {\em MSR}, pages 596--607, 2022.

\bibitem{horwitz1990interprocedural}
Susan Horwitz, Thomas Reps, and David Binkley.
\newblock Interprocedural slicing using dependence graphs.
\newblock {\em TOPLAS}, 12(1):26--60, 1990.

\bibitem{huang2019using}
Zhen Huang, David Lie, Gang Tan, and Trent Jaeger.
\newblock Using safety properties to generate vulnerability patches.
\newblock In {\em S\&P}, pages 539--554, 2019.

\bibitem{iannone2022secret}
Emanuele Iannone, Roberta Guadagni, Filomena Ferrucci, Andrea De~Lucia, and Fabio Palomba.
\newblock The secret life of software vulnerabilities: A large-scale empirical study.
\newblock {\em TSE}, 49(1):44--63, 2022.

\bibitem{vulcost23}
{Ilan Peleg}.
\newblock The high cost of security vulnerabilities.
\newblock \url{https://www.forbes.com/sites/forbesbusinesscouncil /2023/04/10/the-high-cost-of-security-vulnerabilities-why- observability-is-the-solution}, 2023.

\bibitem{cvedashboard23}
{Information Technology Laboratory at NIST}.
\newblock National vulnerability database dashboard.
\newblock \url{https://nvd.nist.gov/general/nvd-dashboard}, 2023.

\bibitem{li2024enhancing}
Haonan Li, Yu~Hao, Yizhuo Zhai, and Zhiyun Qian.
\newblock Enhancing static analysis for practical bug detection: An {LLM}-integrated approach.
\newblock {\em OOPSLA}, 8:474--499, 2024.

\bibitem{li2023starcoder}
Raymond Li, Yangtian Zi, Niklas Muennighoff, Denis Kocetkov, Chenghao Mou, Marc Marone, Christopher Akiki, LI~Jia, Jenny Chim, Qian Liu, et~al.
\newblock {StarCoder}: {May} the source be with you!
\newblock {\em TMLR}.

\bibitem{wen22usenixsecurity}
Wen Li, Jiang Ming, Xiapu Luo, and Haipeng Cai.
\newblock {PolyCruise}: A cross-language dynamic information flow analysis.
\newblock In {\em USENIX Security Symposium}, pages 2513--2530, 2022.

\bibitem{wen23usenixsecurity}
Wen Li, Jinyang Ruan, Guangbei Yi, Long Cheng, Xiapu Luo, and Haipeng Cai.
\newblock {PolyFuzz}: Holistic greybox fuzzing of multi-language systems.
\newblock In {\em USENIX Security Symposium}, pages 1379--1396, 2023.

\bibitem{wen23ccs}
Wen Li, Haoran Yang, Xiapu Luo, Long Cheng, and Haipeng Cai.
\newblock {PyRTFuzz}: Detecting bugs in python runtimes via two-level collaborative fuzzing.
\newblock In {\em CCS}, pages 1645--1659, 2023.

\bibitem{li2021sysevr}
Zhen Li, Deqing Zou, Shouhuai Xu, Hai Jin, Yawei Zhu, and Zhaoxuan Chen.
\newblock {SySeVR}: A framework for using deep learning to detect software vulnerabilities.
\newblock {\em TDSC}, 19(4):2244--2258, 2021.

\bibitem{ma2017vurle}
Siqi Ma, Ferdian Thung, David Lo, Cong Sun, and Robert~H Deng.
\newblock Vurle: Automatic vulnerability detection and repair by learning from examples.
\newblock In {\em ESORICS}, pages 229--246, 2017.

\bibitem{mirskyvulchecker}
Yisroel Mirsky, George Macon, Michael Brown, Carter Yagemann, Matthew Pruett, Evan Downing, Sukarno Mertoguno, and Wenke Lee.
\newblock {VulChecker}: Graph-based vulnerability localization in source code.
\newblock In {\em USENIX Security}, pages 6557--6574, 2023.

\bibitem{noever2023can}
David Noever.
\newblock Can large language models find and fix vulnerable software?
\newblock {\em arXiv preprint arXiv:2308.10345}, 2023.

\bibitem{nong2024chain}
Yu~Nong, Mohammed Aldeen, Long Cheng, Hongxin Hu, Feng Chen, and Haipeng Cai.
\newblock Chain-of-thought prompting of large language models for discovering and fixing software vulnerabilities.
\newblock {\em arXiv preprint arXiv:2402.17230}, 2024.

\bibitem{nongvgx}
Yu~Nong, Richard Fang, Guangbei Yi, Kunsong Zhao, Xiapu Luo, Feng Chen, and Haipeng Cai.
\newblock {VGX}: Large-scale sample generation for boosting learning-based software vulnerability analyses.
\newblock In {\em ICSE}, pages 1--13, 2024.

\bibitem{nong2022generating}
Yu~Nong, Yuzhe Ou, Michael Pradel, Feng Chen, and Haipeng Cai.
\newblock Generating realistic vulnerabilities via neural code editing: {An} empirical study.
\newblock In {\em ESEC/FSE}, pages 1097--1109, 2022.

\bibitem{nongvulgen}
Yu~Nong, Yuzhe Ou, Michael Pradel, Feng Chen, and Haipeng Cai.
\newblock {VulGen}: Realistic vulnerable sample generation via pattern mining and deep learning.
\newblock In {\em ICSE}, pages 2527--2539, 2023.

\bibitem{nong2022open}
Yu~Nong, Rainy Sharma, Abdelwahab Hamou-Lhadj, Xiapu Luo, and Haipeng Cai.
\newblock Open science in software engineering: A study on deep learning-based vulnerability detection.
\newblock {\em TSE}, 49(4):1983--2005, 2022.

\bibitem{nong2024automated}
Yu~Nong, Haoran Yang, Long Cheng, Hongxin Hu, and Haipeng Cai.
\newblock Automated software vulnerability patching using large language models.
\newblock {\em arXiv preprint arXiv:2408.13597}, 2024.

\bibitem{pearce2023examining}
Hammond Pearce, Benjamin Tan, Baleegh Ahmad, Ramesh Karri, and Brendan Dolan-Gavitt.
\newblock Examining zero-shot vulnerability repair with large language models.
\newblock In {\em S\&P}, pages 2339--2356, 2023.

\bibitem{purba2023software}
Moumita~Das Purba, Arpita Ghosh, Benjamin~J Radford, and Bill Chu.
\newblock Software vulnerability detection using large language models.
\newblock In {\em ISSREW}, pages 112--119, 2023.

\bibitem{team2023gemini}
Gemini Team, Rohan Anil, Sebastian Borgeaud, Yonghui Wu, Jean-Baptiste Alayrac, Jiahui Yu, Radu Soricut, Johan Schalkwyk, Andrew~M Dai, Anja Hauth, et~al.
\newblock Gemini: {A} family of highly capable multimodal models.
\newblock {\em arXiv preprint arXiv:2312.11805}, 2023.

\bibitem{touvron2023llama}
Hugo Touvron, Louis Martin, Kevin Stone, Peter Albert, Amjad Almahairi, Yasmine Babaei, Nikolay Bashlykov, Soumya Batra, Prajjwal Bhargava, Shruti Bhosale, et~al.
\newblock Llama 2: Open foundation and fine-tuned chat models.
\newblock {\em arXiv preprint arXiv:2307.09288}, 2023.

\bibitem{ullah2024llms}
Saad Ullah, Mingji Han, Saurabh Pujar, Hammond Pearce, Ayse Coskun, and Gianluca Stringhini.
\newblock {LLMs} cannot reliably identify and reason about security vulnerabilities (yet?): A comprehensive evaluation, framework, and benchmarks.
\newblock In {\em S\&P}, pages 199--199, 2024.

\bibitem{vaswani2017attention}
Ashish Vaswani, Noam Shazeer, Niki Parmar, Jakob Uszkoreit, Llion Jones, Aidan~N Gomez, {\L}ukasz Kaiser, and Illia Polosukhin.
\newblock Attention is all you need.
\newblock {\em NeurIPS}, 30, 2017.

\bibitem{wang2021patchdb}
Xinda Wang, Shu Wang, Pengbin Feng, Kun Sun, and Sushil Jajodia.
\newblock {PatchDB}: A large-scale security patch dataset.
\newblock In {\em DSN}, pages 149--160, 2021.

\bibitem{wang2021codet5}
Yue Wang, Weishi Wang, Shafiq Joty, and Steven~CH Hoi.
\newblock {CodeT5}: Identifier-aware unified pre-trained encoder-decoder models for code understanding and generation.
\newblock In {\em EMNLP}, pages 8696--8708, 2021.

\bibitem{wei2022chain}
Jason Wei, Xuezhi Wang, Dale Schuurmans, Maarten Bosma, Fei Xia, Ed~Chi, Quoc~V Le, Denny Zhou, et~al.
\newblock Chain-of-thought prompting elicits reasoning in large language models.
\newblock {\em NeurIPS}, 35:24824--24837, 2022.

\bibitem{wu2023effective}
Yi~Wu, Nan Jiang, Hung~Viet Pham, Thibaud Lutellier, Jordan Davis, Lin Tan, Petr Babkin, and Sameena Shah.
\newblock How effective are neural networks for fixing security vulnerabilities.
\newblock In {\em ISSTA}, pages 1282--1294, 2023.

\bibitem{joern}
Fabian Yamaguchi.
\newblock A platform for robust analysis of {C/C++} code.
\newblock \url{https://joern.readthedocs.io/en/latest/installation.html}, 2022.

\bibitem{zhang2022example}
Ying Zhang, Ya~Xiao, Md~Mahir~Asef Kabir, Danfeng Yao, and Na~Meng.
\newblock Example-based vulnerability detection and repair in {Java} code.
\newblock In {\em ICPC}, pages 190--201, 2022.

\bibitem{zhang2022program}
Yuntong Zhang, Xiang Gao, Gregory~J Duck, and Abhik Roychoudhury.
\newblock Program vulnerability repair via inductive inference.
\newblock In {\em ISSTA}, pages 691--702, 2022.

\bibitem{zhou2024largeb}
Xin Zhou, Sicong Cao, Xiaobing Sun, and David Lo.
\newblock Large language model for vulnerability detection and repair: Literature review and roadmap.
\newblock {\em TOSEM}, 2024.

\bibitem{zhu2024deepseek}
Qihao Zhu, Daya Guo, Zhihong Shao, Dejian Yang, Peiyi Wang, Runxin Xu, Y~Wu, Yukun Li, Huazuo Gao, Shirong Ma, et~al.
\newblock {DeepSeek-Coder-V2}: Breaking the barrier of closed-source models in code intelligence.
\newblock {\em arXiv preprint arXiv:2406.11931}, 2024.

\bibitem{zxhang2023falcon}
Yoshua~X ZXhang, Yann~M Haxo, and Ying~X Mat.
\newblock {Falcon LLM}: A new frontier in natural language processing.
\newblock {\em AC Investment Research Journal}, 220(44), 2023.

\end{thebibliography}
\bibliographystyle{plain}
}

\appendices

\section{Compostion and Statistics of the Datasets}\label{append:dataset}
Table~\ref{tab:detailed-datasets} shows the composition and statistics of the two datasets. The Zero-Day dataset covers 18 projects and six CWEs where the average number of patched lines is 7.87. The ExtractFix dataset covers seven projects with the same CWEs, while the average number of patched lines is 3.05.

\begin{table*}[h]
  \centering
     \caption{Composition and statistics of the datasets}
     \vspace{2pt}
  \scalebox{0.75}{
\begin{tabular}{|l|l|l|r|}
\hline
\textbf{Dataset} & \textbf{Project-distribution (project:\#CVEs)} & \textbf{CWE-distribution(CWE:\#CVEs)} & \multicolumn{1}{l|}{\textbf{Complexity}} \\
\hline
\multirow{3}[2]{*}{Zero-Day} & linux:73, git:2, ffmpeg:1, less:1, jasper:2, FreeRDP:3, pytorch:1,  Fast-DDS:1,  & CWE787:17, CWE125:23, CWE190:11,  & \multicolumn{1}{l|}{average \#patched lines:7.87} \\
      & esp-idf:1, openssl:1, suricata:1,  RedisBloom:1, tcpdump:1, sngrep:1 & CWE401:11, CWE457:10, CWE476:25	 &  \\
      & jerryscript:2, mynewt-nimble:1, dcmtk:3, libhtp:1 &       &  \\
\hline
\multirow{2}[2]{*}{ExtractFix} & \multirow{2}[2]{*}{coreutils:2, libjpeg:4, libxml2:2, jasper:1, libtiff:8,  binutils-gdb:2, FFmpeg:1} & CWE-787:6, CWE-125:7, CWE-190:5,  & \multicolumn{1}{l|}{average \#patched lines:3.05} \\
      &       & CWE-401:0, CWE-457:1, CWE-476:1 &  \\
\hline
\end{tabular}
}
\label{tab:detailed-datasets}
\end{table*}

\section{Code LLMs against the Zero-Day Dataset}\label{append:codellm}
Table~\ref{tab:detailed-codellm} shows {\tech} effectiveness on code LLMs against the Zero-Day dataset. For each of the models, we use its most powerful version (with most \#paramters)--CodeLlama-70b, CodeQwen-1.5-7b, and DeepSeek-Coder-V2-236b. As shown, the three popular used code LLMs only achieve up to 10.12\% F1 score, much lower than the general-purpose LLMs.

\section{Correct Reasoning Rate Comparison}\label{append:reasoning}
Table~\ref{tab:reasoning} compares the performance of four language models (GPT-4, Gemini-1.5, Claude-3.5, and Llama-3.1) on the Zero-Day dataset using three different prompting approaches. Across all models, {\tech}'s \textit{Vulnerability Semantics Reasoning} approach (highlighted in green) consistently achieves the highest correct reasoning rates. The \emph{Direct Reasoning} and \emph{No Slicing} approaches generally yielded lower success rates, demonstrating the effectiveness of the Vulnerability Semantics Reasoning method.

\begin{table*}[h]
  \centering
     \caption{Code LLMs with {\tech} against the Zero-Day dataset}
     \vspace{2pt}
  \scalebox{0.75}{

\begin{tabular}{|c|r|r|r|r|r|r|r|r|r|r|r|r|}
\hline
\multirow{2}[4]{*}{\textbf{Model}} & \multicolumn{3}{c|}{\textbf{SynEq}} & \multicolumn{3}{c|}{\textbf{SemEq}} & \multicolumn{3}{c|}{\textbf{Plausible}} & \multicolumn{3}{c|}{\textbf{Correct}} \\
\cline{2-13}      & \multicolumn{1}{l|}{\textbf{Recall}} & \multicolumn{1}{l|}{\textbf{Prec}} & \multicolumn{1}{l|}{\textbf{F1}} & \multicolumn{1}{l|}{\textbf{Recall}} & \multicolumn{1}{l|}{\textbf{Prec}} & \multicolumn{1}{l|}{\textbf{F1}} & \multicolumn{1}{l|}{\textbf{Recall}} & \multicolumn{1}{l|}{\textbf{Prec}} & \multicolumn{1}{l|}{\textbf{F1}} & \multicolumn{1}{l|}{\textbf{Recall}} & \multicolumn{1}{l|}{\textbf{Prec}} & \multicolumn{1}{l|}{\textbf{F1}} \\
\hline
CodeLlama & 0.00\% & 0.00\% & \textit{\textbf{0.00\%}} & 1.03\% & 0.42\% & \textit{\textbf{0.60\%}} & 1.03\% & 0.43\% & \textit{\textbf{0.60\%}} & 2.06\% & 0.85\% & \textit{\textbf{1.20\%}} \\
\hline
CodeQwen-1.5 & 0.00\% & 0.00\% & \textit{\textbf{0.00\%}} & 4.12\% & 2.03\% & \textit{\textbf{2.72\%}} & 15.46\% & 4.93\% & \textit{\textbf{7.47\%}} & 18.55\% & 6.95\% & \textit{\textbf{10.12\%}} \\
\hline
DeepSeek-Coder-V2 & 2.06\% & 0.68\% & \textit{\textbf{1.02\%}} & 9.28\% & 3.73\% & \textit{\textbf{5.32\%}} & 6.19\% & 2.03\% & \textit{\textbf{3.06\%}} & 15.46\% & 6.44\% & \textit{\textbf{9.09\%}} \\
\hline
\end{tabular}%

}
\label{tab:detailed-codellm}
\vspace{10pt}
\end{table*}

\begin{table}[h]
  \centering
     \caption{Correct reasoning rate on the Zero-Day dataset}
  \scalebox{0.7}{
\begin{tabular}{|c|l|r|}
\hline
\multicolumn{1}{|l|}{Model} & Prompt & \multicolumn{1}{l|}{Correct Rate} \\
\hline
\multirow{3}[2]{*}{GPT-4} & \cellcolor[rgb]{ .776,  .878,  .706}Vulnerability Semantics Reasoning & \cellcolor[rgb]{ .776,  .878,  .706}\textbf{82.64\%} \\
      & Direct Reasoning & 73.12\% \\
      & No slicing & 79.41\% \\
\hline
\multirow{3}[2]{*}{Gemini-1.5} & \cellcolor[rgb]{ .776,  .878,  .706}Vulnerability Semantics Reasoning & \cellcolor[rgb]{ .776,  .878,  .706}\textbf{63.41\%} \\
      & Direct Reasoning & 59.62\% \\
      & No Slicing & 51.97\% \\
\hline
\multirow{3}[2]{*}{Claude-3.5} & \cellcolor[rgb]{ .776,  .878,  .706}Vulnerability Semantics Reasoning & \cellcolor[rgb]{ .776,  .878,  .706}\textbf{82.64\%} \\
      & Direct Reasoning & 74.18\% \\
      & No Slicing & 79.41\% \\
\hline
\multirow{3}[2]{*}{Llama-3.1} & \cellcolor[rgb]{ .776,  .878,  .706}Vulnerability Semantics Reasoning & \cellcolor[rgb]{ .776,  .878,  .706}\textbf{69.94\%} \\
      & Direct Reasoning & 50.13\% \\
      & No Slicing & 64.74\% \\
\hline
\end{tabular}%

}
\label{tab:reasoning}
\end{table}

\section{More Results on Interprocedural Samples}\label{append:interprocedural}

Table~\ref{tab:detailed-interprocedural} shows the detailed results of {\tech} against the interprocedural samples from the Zero-Day dataest, with \textcolor{red}{red} numbers indicating the best F1 scores achieved by the ablated prompting approaches out of {\tech}. As shown, Claude-3.5 and GPT-4 achieve the best F1 scores (33.57\% and 31.37\%) with {\tech}, outperforming the ablated versions. Specifically, Standard Prompting and Zero-shot Completion have the worst performance (1.79\%-9.81\% and 0.00\%-8.39\% respectively), showing the effectiveness of {\tech} on the interprocedural samples.

\section{Data Leakage Assessment}\label{append:leakage}
We checked ExtractFix dataset against leakage with following LLM queries: 
\begin{enumerate}
    \item \textit{Please tell the Project, CWE-ID, and CVE description of <CVE-ID>.}
    \item \textit{Please tell the CWE-ID and CVE description of <CVE-ID> in <Project>.}
    \item \textit{Please tell the patch of <CVE-ID> which has a <CWE-ID> vulnerability in <Project>.}
\end{enumerate}
The results are shown in Table~\ref{tab:leakage}. As shown, the LLMs either do not know the answers or provide incorrect answers in most of the cases. This indicates that the data leakage issue is not serious.
However, LLMs are largely a black box as we used them in {\tech}, making accurate data leakage assessment very challenging. Thus, we acknowledge that, despite our effort trying to assess the data leakage issues with this dataset, we can not ensure that there was no leakage, nor can we claim that our quantification of the leakage is perfectly accurate. Instead, we took our best effort to assess the leakage, using the queries designed above, as justified as follows. 
First, we consider that there are five important pieces of the knowledge about how a vulnerability is patched (i.e., the knowledge we are concerned that LLM may already have, when we evaluate {\tech} using the ExtractFix dataset which consists of program samples for which these five pieces of information are provided): 
(1) \textit{CVE ID}, (2) \textit{project name}, (3) \textit{CVE description}, (4) \textit{CWE ID}, and (5) \textit{the patch}. 
Then, we design three queries that assess whether an LLM has one or more pieces of such knowledge when given the other pieces. More specifically:
\begin{itemize}
\item \textbf{Query 1}: we provide the LLM with (1) and ask it about (2), (3), and (4); 
\item \textbf{Query 2}: we provide the LLM with (1) and (2), and ask it about (3) and (4); 
\item \textbf{Query 3}: we provide the LLM with (1), (2), and (4), and ask it about (5). 
\end{itemize}

As seen, with these three queries sent to the LLM in order of Queries 1, 2, 3, we gradually provide the LLM with increasingly more pieces of information, and check if the LLM knows about the rest. This helps us check the LLM against the data leakage with varying levels of difficulty, and put together we believe the LLM’s responses can give us a perspective into the data leakage issue. 
Nevertheless, we could not make bold claims that these queries are the best to assess data leakage about vulnerability patches in general. 
Moreover, we would like to note the following points: 
(1) as a comparison, we also test the LLMs with a well-known vulnerability CVE-2021-44228 using the same three queries. The LLMs did successfully answer some of five pieces of information about this CVE correctly, indicating that our queries are helpful to test the data leakage.
(2) note that {\tech}’s performance evaluation is primarily based on the Zero-Day dataset, and we have compared it with both LLM-based and non-LLM-based baseline patching techniques on this dataset. The reason we use the ExtractFix dataset is mainly for being able to compare with two additional state-of-the-art baselines (ExtractFix~\cite{gao2021beyond} and VulnFix~\cite{zhang2022program}).

\begin{table}[h]
  \centering
     \caption{Data leakage test results on the ExtractFix dataset with the format “model-does-not-know\%, incorrect-answer\%, correct-answer\%”}
     \vspace{4pt}
  \scalebox{0.7}{
\begin{tabular}{|r|l|l|l|l|}
\hline
\multicolumn{1}{|l|}{\textbf{Query}} & \textbf{GPT-4} & \textbf{Gemini-1.5} & \textbf{Claude-3.5} & \textbf{Llama-3.1} \\
\hline
1     & 100\%, 0\%, 0\% & 0\%, 100\%, 0\% & 10\%, 90\%, 0\% & 25\%, 70\%, 5\% \\
\hline
2     & 30\%, 60\%, 10\% & 5\%, 85\%, 10\% & 55\%, 45\%, 0\% & 5\%, 90\%, 5\% \\
\hline
3     & 100\%, 0\%, 0\% & 25\%, 75\%, 0\% & 100\%, 0\%, 0\% & 70\%, 30\%, 0\% \\
\hline
\end{tabular}%

}
\label{tab:leakage}
\end{table}

\section{More End-to-end Experiment Results}\label{append:usability}

Table~\ref{tab:detailed-codeql} shows the detailed end-to-end integration results with CodeQL against the Zero-Day dataset. As shown, {\tech} achieves 23.85\% F1 with fully automated method and 27.29\% with realistic method. In contrast, the baselines approaches and models have much less effectiveness, indicating that {\tech} can be realistically used in practice.

\vspace{20pt}
\section{More Details on Case Study Examples}\label{append:task2}

\begin{figure}[h]
\vspace{2pt}
\centering
	\includegraphics[width=1\linewidth]{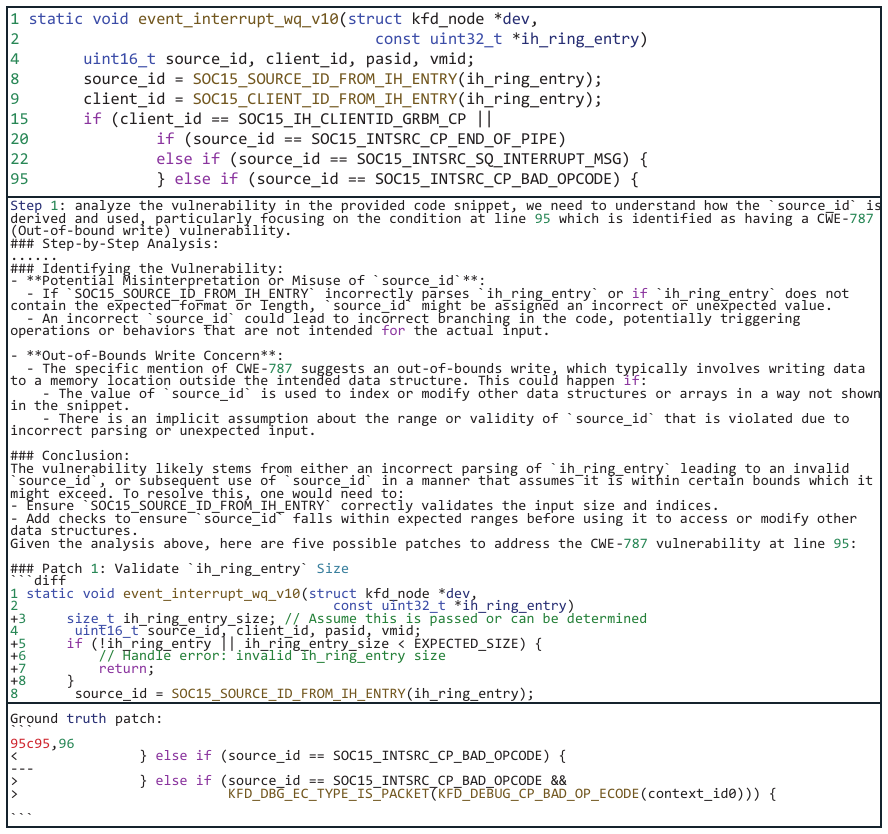}
        \vspace{-12pt}
        \caption{A failure case where the vulnerability is incorrectly identified.}
	\label{fig:failure-1}
        \vspace{4pt}
\end{figure}

\begin{table*}[h]
  \centering
     \caption{Detailed results of {\tech} against the interprocedural samples from the Zero-Day dataset}
     \vspace{2pt}
  \scalebox{0.6}{

\begin{tabular}{|c|l|r|r|r|r|r|r|r|r|r|r|r|r|}
\hline
\multirow{3}[0]{*}{Model} & \multicolumn{1}{c|}{\multirow{3}[0]{*}{Approach}} & \multicolumn{12}{c|}{Zero-Day Dataset} \\
\cline{3-14}      &       & \multicolumn{3}{c|}{\textbf{SynEq}} & \multicolumn{3}{c|}{\textbf{SemEq}} & \multicolumn{3}{c|}{\textbf{Plausible}} & \multicolumn{3}{c|}{\textbf{Correct}} \\
\cline{3-14}      &       & \multicolumn{1}{l|}{\textbf{Recall}} & \multicolumn{1}{l|}{\textbf{Prec}} & \multicolumn{1}{l|}{\textbf{F1}} & \multicolumn{1}{l|}{\textbf{Recall}} & \multicolumn{1}{l|}{\textbf{Prec}} & \multicolumn{1}{l|}{\textbf{F1}} & \multicolumn{1}{l|}{\textbf{Recall}} & \multicolumn{1}{l|}{\textbf{Prec}} & \multicolumn{1}{l|}{\textbf{F1}} & \multicolumn{1}{l|}{\textbf{Recall}} & \multicolumn{1}{l|}{\textbf{Prec}} & \multicolumn{1}{l|}{\textbf{F1}} \\
\hline
\multirow{8}[2]{*}{GPT-4} & {\tech}  & \cellcolor[rgb]{ .776,  .878,  .706}4.76\% & \cellcolor[rgb]{ .776,  .878,  .706}1.11\% & \cellcolor[rgb]{ .776,  .878,  .706}\textit{\textbf{1.80\%}} & \cellcolor[rgb]{ .776,  .878,  .706}9.52\% & \cellcolor[rgb]{ .776,  .878,  .706}4.44\% & \cellcolor[rgb]{ .776,  .878,  .706}\textit{\textbf{6.06\%}} & \cellcolor[rgb]{ .776,  .878,  .706}38.10\% & \cellcolor[rgb]{ .776,  .878,  .706}21.11\% & \cellcolor[rgb]{ .776,  .878,  .706}\textit{\textbf{27.17\%}} & \cellcolor[rgb]{ .776,  .878,  .706}38.10\% & \cellcolor[rgb]{ .776,  .878,  .706}26.67\% & \cellcolor[rgb]{ .776,  .878,  .706}\textit{\textbf{31.37\%}} \\
      & No Validation & \cellcolor[rgb]{ .886,  .937,  .855}4.76\% & \cellcolor[rgb]{ .886,  .937,  .855}0.90\% & \cellcolor[rgb]{ .886,  .937,  .855}\textit{1.52\%} & \cellcolor[rgb]{ .886,  .937,  .855}14.29\% & \cellcolor[rgb]{ .886,  .937,  .855}5.41\% & \cellcolor[rgb]{ .886,  .937,  .855}\textit{7.84\%} & \cellcolor[rgb]{ .886,  .937,  .855}38.10\% & \cellcolor[rgb]{ .886,  .937,  .855}18.92\% & \cellcolor[rgb]{ .886,  .937,  .855}\textit{25.28\%} & \cellcolor[rgb]{ .886,  .937,  .855}38.10\% & \cellcolor[rgb]{ .886,  .937,  .855}25.23\% & \cellcolor[rgb]{ .886,  .937,  .855}\textit{30.35\%} \\
      & No Slicing & 0.00\% & 0.00\% & \textit{0.00\%} & 9.52\% & 1.82\% & \textit{3.05\%} & 33.33\% & 16.36\% & \textit{21.95\%} & 33.33\% & 18.18\% & \textit{23.53\%} \\
      & Random Exemplars & 4.76\% & 0.90\% & \textit{1.52\%} & 9.52\% & 1.80\% & \textit{3.03\%} & 28.57\% & 12.61\% & \textit{17.50\%} & 33.33\% & 15.32\% & \textit{20.99\%} \\
      & Manual Exemplars & 0.00\% & 0.00\% & \textit{0.00\%} & 9.52\% & 2.73\% & \textit{4.24\%} & 33.33\% & 14.55\% & \textit{20.25\%} & 38.10\% & 17.27\% & \textcolor[rgb]{ 1,  0,  0}{\textit{23.77\%}} \\
      & Direct Reasoning & 0.00\% & 0.00\% & \textit{0.00\%} & 4.76\% & 0.90\% & \textit{1.52\%} & 38.10\% & 12.61\% & \textit{18.95\%} & 38.10\% & 13.51\% & \textit{19.95\%} \\
      & Standard Prompting & 0.00\% & 0.00\% & 0.00\% & 0.00\% & 0.00\% & 0.00\% & 19.05\% & 6.60\% & 9.81\% & 19.05\% & 6.60\% & 9.81\% \\
      & Zero-shot Completion & 0.00\% & 0.00\% & 0.00\% & 0.00\% & 0.00\% & 0.00\% & 9.52\% & 4.72\% & 6.31\% & 9.52\% & 4.72\% & 6.31\% \\
\hline
\multirow{8}[2]{*}{Gemini-1.5} & {\tech}  & \cellcolor[rgb]{ .776,  .878,  .706}0.00\% & \cellcolor[rgb]{ .776,  .878,  .706}0.00\% & \cellcolor[rgb]{ .776,  .878,  .706}\textit{\textbf{0.00\%}} & \cellcolor[rgb]{ .776,  .878,  .706}9.52\% & \cellcolor[rgb]{ .776,  .878,  .706}2.02\% & \cellcolor[rgb]{ .776,  .878,  .706}\textit{\textbf{3.33\%}} & \cellcolor[rgb]{ .776,  .878,  .706}28.57\% & \cellcolor[rgb]{ .776,  .878,  .706}12.12\% & \cellcolor[rgb]{ .776,  .878,  .706}\textit{\textbf{17.02\%}} & \cellcolor[rgb]{ .776,  .878,  .706}28.57\% & \cellcolor[rgb]{ .776,  .878,  .706}14.14\% & \cellcolor[rgb]{ .776,  .878,  .706}\textit{\textbf{18.91\%}} \\
      & No Validation & \cellcolor[rgb]{ .886,  .937,  .855}0.00\% & \cellcolor[rgb]{ .886,  .937,  .855}0.00\% & \cellcolor[rgb]{ .886,  .937,  .855}\textit{0.00\%} & \cellcolor[rgb]{ .886,  .937,  .855}9.52\% & \cellcolor[rgb]{ .886,  .937,  .855}2.02\% & \cellcolor[rgb]{ .886,  .937,  .855}\textit{3.33\%} & \cellcolor[rgb]{ .886,  .937,  .855}28.57\% & \cellcolor[rgb]{ .886,  .937,  .855}11.32\% & \cellcolor[rgb]{ .886,  .937,  .855}\textit{16.22\%} & \cellcolor[rgb]{ .886,  .937,  .855}28.57\% & \cellcolor[rgb]{ .886,  .937,  .855}13.34\% & \cellcolor[rgb]{ .886,  .937,  .855}\textit{18.18\%} \\
      & No Slicing & 0.00\% & 0.00\% & \textit{0.00\%} & 9.52\% & 2.06\% & \textit{3.39\%} & 14.28\% & 10.31\% & \textit{11.98\%} & 23.81\% & 12.37\% & \textcolor[rgb]{ 1,  0,  0}{\textit{16.28\%}} \\
      & Random Exemplars & 0.00\% & 0.00\% & \textit{0.00\%} & 9.52\% & 2.08\% & \textit{3.42\%} & 19.05\% & 9.38\% & \textit{12.57\%} & 23.81\% & 11.46\% & \textit{15.47\%} \\
      & Manual Exemplars & 0.00\% & 0.00\% & \textit{0.00\%} & 9.52\% & 2.02\% & \textit{3.33\%} & 23.81\% & 9.09\% & \textit{13.16\%} & 23.81\% & 11.11\% & \textit{15.15\%} \\
      & Direct Reasoning & 0.00\% & 0.00\% & 0.00\% & 0.00\% & 0.00\% & 0.00\% & 23.81\% & 7.61\% & 11.54\% & 23.81\% & 7.62\% & 11.54\% \\
      & Standard Prompting & 0.00\% & 0.00\% & 0.00\% & 0.00\% & 0.00\% & 0.00\% & 4.76\% & 1.10\% & 1.79\% & 4.76\% & 1.10\% & 1.79\% \\
      & Zero-shot Completion & 0.00\% & 0.00\% & 0.00\% & 0.00\% & 0.00\% & 0.00\% & 0.00\% & 0.00\% & 0.00\% & 0.00\% & 0.00\% & 0.00\% \\
\hline
\multirow{8}[2]{*}{Claude-3.5} & {\tech}  & \cellcolor[rgb]{ .776,  .878,  .706}4.76\% & \cellcolor[rgb]{ .776,  .878,  .706}0.93\% & \cellcolor[rgb]{ .776,  .878,  .706}\textit{\textbf{1.55\%}} & \cellcolor[rgb]{ .776,  .878,  .706}19.05\% & \cellcolor[rgb]{ .776,  .878,  .706}3.70\% & \cellcolor[rgb]{ .776,  .878,  .706}\textit{\textbf{6.20\%}} & \cellcolor[rgb]{ .776,  .878,  .706}47.62\% & \cellcolor[rgb]{ .776,  .878,  .706}21.30\% & \cellcolor[rgb]{ .776,  .878,  .706}\textit{\textbf{29.43\%}} & \cellcolor[rgb]{ .776,  .878,  .706}47.62\% & \cellcolor[rgb]{ .776,  .878,  .706}25.93\% & \cellcolor[rgb]{ .776,  .878,  .706}\textit{\textbf{33.57\%}} \\
      & No Validation & \cellcolor[rgb]{ .886,  .937,  .855}4.76\% & \cellcolor[rgb]{ .886,  .937,  .855}0.90\% & \cellcolor[rgb]{ .886,  .937,  .855}\textit{1.52\%} & \cellcolor[rgb]{ .886,  .937,  .855}19.05\% & \cellcolor[rgb]{ .886,  .937,  .855}3.60\% & \cellcolor[rgb]{ .886,  .937,  .855}\textit{6.06\%} & \cellcolor[rgb]{ .886,  .937,  .855}47.62\% & \cellcolor[rgb]{ .886,  .937,  .855}20.72\% & \cellcolor[rgb]{ .886,  .937,  .855}\textit{28.88\%} & \cellcolor[rgb]{ .886,  .937,  .855}47.62\% & \cellcolor[rgb]{ .886,  .937,  .855}25.23\% & \cellcolor[rgb]{ .886,  .937,  .855}\textit{32.98\%} \\
      & No Slicing & 0.00\% & 0.00\% & \textit{0.00\%} & 9.52\% & 1.98\% & \textit{3.28\%} & 38.10\% & 19.80\% & \textit{26.06\%} & 38.10\% & 21.78\% & \textit{27.72\%} \\
      & Random Exemplars & 0.00\% & 0.00\% & \textit{0.00\%} & 4.76\% & 0.90\% & \textit{1.52\%} & 42.86\% & 18.92\% & \textit{26.25\%} & 47.62\% & 19.82\% & \textcolor[rgb]{ 1,  0,  0}{\textit{27.99\%}} \\
      & Manual Exemplars & 0.00\% & 0.00\% & \textit{0.00\%} & 14.29\% & 2.70\% & \textit{4.55\%} & 42.86\% & 16.22\% & \textit{23.53\%} & 42.86\% & 18.92\% & \textit{26.25\%} \\
      & Direct Reasoning & 0.00\% & 0.00\% & 0.00\% & 9.52\% & 1.89\% & 3.15\% & 38.10\% & 17.92\% & 24.38\% & 42.86\% & 19.81\% & 27.10\% \\
      & Standard Prompting & 0.00\% & 0.00\% & 0.00\% & 4.76\% & 1.96\% & 2.78\% & 9.52\% & 4.90\% & 6.47\% & 14.29\% & 6.86\% & 9.27\% \\
      & Zero-shot Completion & 0.00\% & 0.00\% & 0.00\% & 0.00\% & 0.00\% & 0.00\% & 14.29\% & 5.94\% & 8.39\% & 14.29\% & 5.94\% & 8.39\% \\
\hline
\multirow{8}[2]{*}{Llama-3.1} & {\tech}  & \cellcolor[rgb]{ .776,  .878,  .706}0.00\% & \cellcolor[rgb]{ .776,  .878,  .706}0.00\% & \cellcolor[rgb]{ .776,  .878,  .706}\textit{\textbf{0.00\%}} & \cellcolor[rgb]{ .776,  .878,  .706}9.52\% & \cellcolor[rgb]{ .776,  .878,  .706}2.35\% & \cellcolor[rgb]{ .776,  .878,  .706}\textit{\textbf{3.77\%}} & \cellcolor[rgb]{ .776,  .878,  .706}38.10\% & \cellcolor[rgb]{ .776,  .878,  .706}15.29\% & \cellcolor[rgb]{ .776,  .878,  .706}\textit{\textbf{21.82\%}} & \cellcolor[rgb]{ .776,  .878,  .706}38.10\% & \cellcolor[rgb]{ .776,  .878,  .706}17.65\% & \cellcolor[rgb]{ .776,  .878,  .706}\textit{\textbf{24.12\%}} \\
      & No Validation & \cellcolor[rgb]{ .886,  .937,  .855}0.00\% & \cellcolor[rgb]{ .886,  .937,  .855}0.00\% & \cellcolor[rgb]{ .886,  .937,  .855}\textit{0.00\%} & \cellcolor[rgb]{ .886,  .937,  .855}9.52\% & \cellcolor[rgb]{ .886,  .937,  .855}1.80\% & \cellcolor[rgb]{ .886,  .937,  .855}\textit{3.03\%} & \cellcolor[rgb]{ .886,  .937,  .855}38.10\% & \cellcolor[rgb]{ .886,  .937,  .855}12.61\% & \cellcolor[rgb]{ .886,  .937,  .855}\textit{18.95\%} & \cellcolor[rgb]{ .886,  .937,  .855}38.10\% & \cellcolor[rgb]{ .886,  .937,  .855}14.41\% & \cellcolor[rgb]{ .886,  .937,  .855}\textit{20.91\%} \\
      & No Slicing & 0.00\% & 0.00\% & \textit{0.00\%} & 4.76\% & 3.85\% & \textit{4.26\%} & 9.52\% & 19.23\% & \textit{12.74\%} & 14.29\% & 23.08\% & \textcolor[rgb]{ 1,  0,  0}{\textit{17.65\%}} \\
      & Random Exemplars & 0.00\% & 0.00\% & \textit{0.00\%} & 4.76\% & 1.98\% & \textit{2.80\%} & 19.05\% & 8.91\% & \textit{12.14\%} & 23.81\% & 10.89\% & \textit{14.95\%} \\
      & Manual Exemplars & 0.00\% & 0.00\% & \textit{0.00\%} & 4.76\% & 2.56\% & \textit{3.33\%} & 19.05\% & 12.82\% & \textit{15.33\%} & 19.05\% & 15.38\% & \textit{17.02\%} \\
      & Direct Reasoning & \textit{0.00\%} & \textit{0.00\%} & \textit{0.00\%} & \textit{0.00\%} & \textit{0.00\%} & \textit{0.00\%} & \textit{28.57\%} & \textit{10.81\%} & \textit{15.68\%} & \textit{28.57\%} & \textit{10.81\%} & \textit{15.68\%} \\
      & Standard Prompting & \textit{0.00\%} & \textit{0.00\%} & \textit{0.00\%} & \textit{0.00\%} & \textit{0.00\%} & \textit{0.00\%} & \textit{9.52\%} & \textit{3.13\%} & \textit{4.71\%} & \textit{9.52\%} & \textit{3.13\%} & \textit{4.71\%} \\
      & Zero-shot Completion & \textit{0.00\%} & \textit{0.00\%} & \textit{0.00\%} & \textit{0.00\%} & \textit{0.00\%} & \textit{0.00\%} & \textit{9.52\%} & \textit{2.70\%} & \textit{4.21\%} & \textit{9.52\%} & \textit{2.70\%} & \textit{4.21\%} \\
\hline
\end{tabular}
}
\label{tab:detailed-interprocedural}
\end{table*}

\begin{table*}[h]
  \centering
     \caption{Detailed end-to-end results with vulnerability analyzer CodeQL against the Zero-Day dataset}
     \vspace{2pt}
  \scalebox{0.6}{

\begin{tabular}{|c|c|c|r|r|r|r|r|r|r|r|r|r|r|r|}
\hline
\multirow{3}[0]{*}{Method} & \multirow{3}[0]{*}{Model} & \multirow{3}[0]{*}{Approach} & \multicolumn{12}{c|}{Zero-Day Dataset} \\
\cline{4-15}      &       &       & \multicolumn{3}{c|}{\textbf{SynEq}} & \multicolumn{3}{c|}{\textbf{SemEq}} & \multicolumn{3}{c|}{\textbf{Plausible}} & \multicolumn{3}{c|}{\textbf{Correct}} \\
\cline{4-15}      &       &       & \multicolumn{1}{l|}{\textbf{Recall}} & \multicolumn{1}{l|}{\textbf{Prec}} & \multicolumn{1}{l|}{\textbf{F1}} & \multicolumn{1}{l|}{\textbf{Recall}} & \multicolumn{1}{l|}{\textbf{Prec}} & \multicolumn{1}{l|}{\textbf{F1}} & \multicolumn{1}{l|}{\textbf{Recall}} & \multicolumn{1}{l|}{\textbf{Prec}} & \multicolumn{1}{l|}{\textbf{F1}} & \multicolumn{1}{l|}{\textbf{Recall}} & \multicolumn{1}{l|}{\textbf{Prec}} & \multicolumn{1}{l|}{\textbf{F1}} \\
\hline
\multirow{6}[0]{*}{Fully Automated} & \multirow{4}[2]{*}{Claude-3.5} & \multicolumn{1}{l|}{{\tech}} & \cellcolor[rgb]{ .776,  .878,  .706}1.03\% & \cellcolor[rgb]{ .776,  .878,  .706}0.33\% & \cellcolor[rgb]{ .776,  .878,  .706}\textit{\textbf{0.50\%}} & \cellcolor[rgb]{ .776,  .878,  .706}12.37\% & \cellcolor[rgb]{ .776,  .878,  .706}7.21\% & \cellcolor[rgb]{ .776,  .878,  .706}\textit{\textbf{9.11\%}} & \cellcolor[rgb]{ .776,  .878,  .706}26.80\% & \cellcolor[rgb]{ .776,  .878,  .706}10.82\% & \cellcolor[rgb]{ .776,  .878,  .706}\textit{\textbf{15.42\%}} & \cellcolor[rgb]{ .776,  .878,  .706}34.02\% & \cellcolor[rgb]{ .776,  .878,  .706}18.36\% & \cellcolor[rgb]{ .776,  .878,  .706}\textit{\textbf{23.85\%}} \\
      &       & \multicolumn{1}{l|}{Manual Exemplars} & 1.03\% & 0.32\% & \textit{0.49\%} & 7.22\% & 5.18\% & \textit{6.03\%} & 15.46\% & 5.83\% & \textit{8.46\%} & 22.68\% & 11.33\% & \textit{15.11\%} \\
      &       & \multicolumn{1}{l|}{Standard Prompting} & 0.00\% & 0.00\% & 0.00\% & 8.25\% & 5.25\% & 6.41\% & 11.34\% & 5.90\% & 7.76\% & 16.49\% & 11.15\% & 13.30\% \\
      &       & \multicolumn{1}{l|}{Zero-shot Completion} & 0.00\% & 0.00\% & 0.00\% & 2.06\% & 0.78\% & 1.13\% & 14.43\% & 8.27\% & 10.51\% & 15.46\% & 9.04\% & 11.41\% \\
\cline{2-15}      & VulRepair & -     & 2.06\% & 0.47\% & 0.77\% & 13.40\% & 8.47\% & 10.38\% & 1.03\% & 0.24\% & 0.38\% & 16.49\% & 9.18\% & 11.79\% \\
\cline{2-15}      & Getafix & -     & 5.15\% & 5.26\% & 5.21\% & 0.00\% & 0.00\% & 0.00\% & 1.03\% & 0.44\% & 0.62\% & 6.19\% & 5.70\% & 5.93\% \\
\hline
\multirow{6}[0]{*}{Realistic} & \multirow{4}[2]{*}{Claude-3.5} & \multicolumn{1}{l|}{{\tech}} & \cellcolor[rgb]{ .776,  .878,  .706}1.03\% & \cellcolor[rgb]{ .776,  .878,  .706}0.25\% & \cellcolor[rgb]{ .776,  .878,  .706}\textit{\textbf{0.40\%}} & \cellcolor[rgb]{ .776,  .878,  .706}19.59\% & \cellcolor[rgb]{ .776,  .878,  .706}7.71\% & \cellcolor[rgb]{ .776,  .878,  .706}\textit{\textbf{11.07\%}} & \cellcolor[rgb]{ .776,  .878,  .706}29.90\% & \cellcolor[rgb]{ .776,  .878,  .706}12.44\% & \cellcolor[rgb]{ .776,  .878,  .706}\textit{\textbf{17.57\%}} & \cellcolor[rgb]{ .776,  .878,  .706}41.24\% & \cellcolor[rgb]{ .776,  .878,  .706}20.40\% & \cellcolor[rgb]{ .776,  .878,  .706}\textit{\textbf{27.29\%}} \\
      &       & \multicolumn{1}{l|}{Manual Exemplars} & 0.00\% & 0.00\% & \textit{0.00\%} & 15.46\% & 6.05\% & \textit{8.70\%} & 15.46\% & 5.33\% & \textit{7.92\%} & 25.77\% & 11.38\% & \textit{15.79\%} \\
      &       & \multicolumn{1}{l|}{Standard Prompting} & 1.03\% & 0.28\% & 0.44\% & 19.59\% & 9.47\% & 12.77\% & 13.40\% & 6.41\% & 8.67\% & 24.74\% & 16.16\% & 19.55\% \\
      &       & \multicolumn{1}{l|}{Zero-shot Completion} & 0.00\% & 0.00\% & 0.00\% & 2.06\% & 0.78\% & 1.13\% & 14.43\% & 8.27\% & 10.51\% & 15.46\% & 9.04\% & 11.41\% \\
\cline{2-15}      & VulRepair & -     & 2.06\% & 0.47\% & 0.77\% & 13.40\% & 8.47\% & 10.38\% & 1.03\% & 0.24\% & 0.38\% & 16.49\% & 9.18\% & 11.79\% \\
\cline{2-15}      & Getafix & -     & 5.15\% & 5.26\% & 5.21\% & 0.00\% & 0.00\% & 0.00\% & 1.03\% & 0.44\% & 0.62\% & 6.19\% & 5.70\% & 5.93\% \\
\hline
\end{tabular}
}
\label{tab:detailed-codeql}
\vspace{10pt}
\end{table*}

We conduct case studies on the {\tech}'s failure cases to investigate the symptoms and root causes. As discussed in Section~\ref{sec:fail}, we show two examples for the failure symptoms and root causes.

Figure~\ref{fig:failure-1} shows an example where the failed patching attempt indicates that the original vulnerability was not accurately identified or addressed. The symptom was incorrectly diagnosed as an out-of-bounds write (CWE-787) related to the `source\_id` variable, leading to proposed patches that focused on validating and constraining this value. However, the root cause appears to be a logic error or insufficient condition checking in the original code. The actual fix involves adding an additional condition to check if the context ID corresponds to a specific packet type, suggesting that the vulnerability was more likely related to improper input validation or incorrect handling of certain packet types, rather than a straightforward buffer overflow issue.

\begin{figure}[h]
\centering
	\includegraphics[width=1\linewidth]{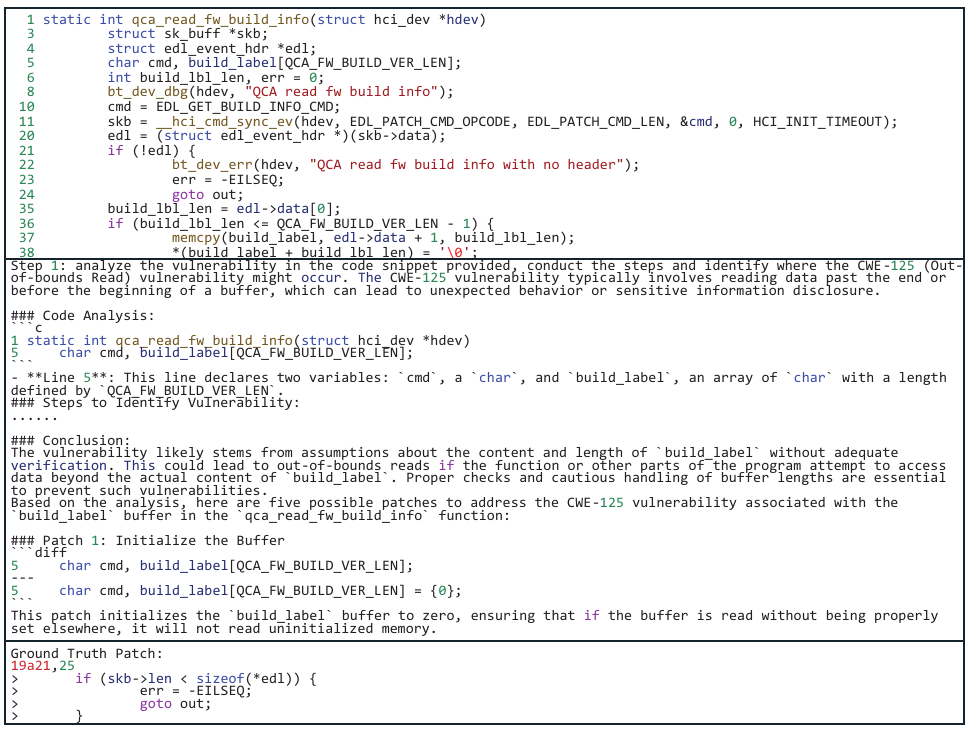}
        \vspace{-15pt}
        \caption{A failure case where the patch has insufficient or incorrect code modification.}
	\label{fig:failure-2}
        \vspace{-2pt}
\end{figure}

Figure~\ref{fig:failure-2} shows an example where failed patching attempts primarily focused on addressing potential issues with the `build\_label` array's initialization and usage, but they missed the root cause of the vulnerability. The actual problem stems from insufficient bounds checking when accessing data from the `skb` (socket buffer) structure. The ground truth patch reveals that the vulnerability is caused by inadequate validation of the `skb->len` against the expected data sizes at various points in the function. This allows for potential out-of-bounds reads when accessing `edl` structure members and data. The correct fix involves adding multiple checks to ensure that the `skb` contains enough data before accessing its contents, and dynamically allocating memory for `build\_label` instead of using a fixed-size array, which addresses both buffer overflow and potential memory leaks.

\end{document}